\documentclass[a4paper,11pt]{article}

\usepackage{blindtext,titlefoot}

\usepackage{amsfonts,amsthm,amssymb,amsmath}
\usepackage[title,titletoc,toc]{appendix}
\usepackage[utf8]{inputenc}
\usepackage[affil-it]{authblk}
\usepackage{xcolor}
\usepackage{tikz}
\newcommand{\RN}[1]{%
  \textup{\uppercase\expandafter{\romannumeral#1}}%
}
\newcommand\COMP{\hbox{C\kern -.58em {\raise .54ex \hbox{$\scriptscriptstyle |$}}
\kern-.55em {\raise .53ex \hbox{$\scriptscriptstyle |$}} }}
\newcommand\NN{\hbox{I\kern-.2em\hbox{N}}}
\newcommand\RR{\hbox{I\kern-.2em\hbox{R}}}
\newcommand\sRR{{\it \hbox{I\kern-.2em\hbox{R}}}}
\newcommand\QQ{\hbox{I\kern-.53em\hbox{Q}}}
\newcommand\PP{\hbox{I\kern-.53em\hbox{P}}}
\newcommand\EE{\hbox{I\kern-.53em\hbox{E}}}
\newcommand\ZZ{{{\rm Z}\kern-.28em{\rm Z}}}
\newcommand\be{\begin{equation}}
\newcommand\ee{\end{equation}}

\usepackage{scalerel,stackengine}
\stackMath
\newcommand\reallywidehat[1]{%
\savestack{\tmpbox}{\stretchto{%
  \scaleto{ \scalerel*[\widthof{\ensuremath{#1}}]{\kern.1pt\mathchar"0362\kern.1pt} {\rule{0ex}{\textheight}} }{\textheight}}{2.4ex}}
\stackon[-6.9pt]{#1}{\tmpbox}}

\newtheorem{theorem}{Theorem}[section]

\newtheorem{proposition}[theorem]{Proposition}
\newtheorem{remark}[theorem]{Remark}

\newtheorem{example}[theorem]{Example}
\newtheorem{lemma}[theorem]{Lemma}

\newtheorem{corollary}[theorem]{Corollary}
\newtheorem{definition}[theorem]{Definition}

\makeatletter
\newcommand*\bigcdot{\mathpalette\bigcdot@{.5}}
\newcommand*\bigcdot@[2]{\mathbin{\vcenter{\hbox{\scalebox{#2}{$\m@th#1\bullet$}}}}}
\makeatother
\newcommand{\is}{\bigcdot }

\def \Lbrack {[\![}
\def \Rbrack {]\!]}

\newcommand\cF{{\cal F}}
\newcommand\cG{{\cal G}}

\newcommand\cH{{\cal H}}

\newcommand\cP{{\cal P}}

\newcommand\R{{\bf R}}

\newcommand\beq{\begin{equation}}
\newcommand\eeq{\end{equation}}
\newcommand\beqa{\begin{equation*}}
\newcommand\eeqa{\end{equation*}}
\newcommand\bea{\begin{eqnarray}}
\newcommand\eea{\end{eqnarray}}
\newcommand\bean{\begin{eqnarray*}}
\newcommand\eean{\end{eqnarray*}}

\DeclareMathOperator{\essinf}{ess\;inf}
\DeclareMathOperator{\esssup}{ess\;sup}

\newcommand\bP{{\bf {\rm P}}}

\numberwithin{equation}{section}

\begin{document}

\title{Super-hedging-pricing formulas and Immediate-Profit arbitrage for market models under random horizon
}

\author{Tahir Choulli\\ 
Mathematical and Statistical Sciences Dept.\\
University of Alberta, Edmonton, Canada
\and 
Emmanuel Lepinette\\ 
Ceremade, UMR  CNRS 7534,  Paris Dauphine University,\\
PSL National Research, Place du Mar\'echal De Lattre De Tassigny, \\
75775 Paris cedex 16, France and\\
Gosaef, Faculty of Sciences of Tunis, Tunisia.\\
Email: emmanuel.lepinette@ceremade.dauphine.fr}

\maketitle

\begin{abstract}
 In this paper, we consider the discrete-time setting, and the market model described by $(S, \mathbb{F},\tau)$. Herein $\mathbb{F}$ is the ``public" flow of information which is available to all agents overtime, $S$ is the discounted price process of $d$-tradable assets, and $\tau$ is an arbitrary random time whose occurrence might not be observable via $\mathbb{F}$. Thus, we consider the larger flow $\mathbb{G}$ which incorporates $\mathbb{F}$ and makes $\tau$ an observable random time. This framework covers the credit risk theory setting, the life insurance setting and the setting of employee stock option valuation. For the stopped model $(S^{\tau},\mathbb{G})$ and for various vulnerable claims, based on this model, we address the super-hedging pricing valuation problem and its intrinsic Immediate-Profit arbitrage (IP hereafter for short). Our first main contribution lies in singling out the impact of change of prior and/or information on {\it conditional essential supremum}, which is a vital tool in super-hedging pricing. The second main contribution consists of describing as explicit as possible how the set of super-hedging prices expands under the stochasticity of $\tau$ and its risks,  and we address the IP arbitrage for  $(S^{\tau},\mathbb{G})$ as well. The third main contribution resides in elaborating as explicit as possible pricing formulas for vulnerable claims, and singling out the various informational risks in the prices' dynamics. 

\end{abstract}

\section{Introduction}
\label{intro}

In this paper, we consider a general {\it{initial}} discrete-time model represented by the pair $(S,\mathbb{F})$ and an arbitrary random horizon $\tau$. Herein, $S$ is the assets' price process and $\mathbb{F}$ is the ``public" flow of information which is available to all agents overtime, while the random time $\tau$ might not be observable via the flow $\mathbb{F}$. This random time represents the default of a firm in credit risk theory, the death time of an insured in life insurance,  the job's termination time of an {\it employee-stock-option}'s holder (called ESO hereafter) in finance, ..., etcetera. Hence, our current setting covers these three aforementioned frameworks, and for more about these we refer the reader to \cite{BelangerShreveWong,BieleckiRutkowski,Szimayer,Sircar,CarrLinetsky,LeungSircar,Zervos} and the references therein to cite a few.  As a random time can not be observable before its occurrence, the  progressive enlargement approach of $\mathbb{F}$ with $\tau$ seems the tailor-fit method for our setting in modelling mathematically the large flow of information which incorporates both $\mathbb{F}$ and $\tau$.  This larger flow will be denoted throughout the paper by $\mathbb{G}$ and will be defined more precisely in the next section. Therefore, our main objective lies in addressing the evaluation problem for the stopped financial model $(S^{\tau},\mathbb{G},P)$, and focus on the super-hedging pricing approach and its intrinsic arbitrage called {\it immediate-profit}.\\

The super-hedging price of the financial claim/asset $C$ is the infinimum amount required to initiate a hedging strategy for $C$. The super-hedging pricing concept was introduced in Bensaid et al. in \cite{Bensaid} for the binomial framework with transaction costs. Afterwards, the characterization and/or computation of the super-hedging price became a central problem in mathematical finance, and we refer the reader to \cite{Follmer,Karatzas-Shreve} and the references therein to cite a few. These studies were carried out under the main assumption of no-arbitrage using the fundamental theorem of asset pricing theorem (FTAP hereafter). Using this FTAP, the main results on the super-hedging pricing consist of establishing a dual formulation for the price using martingale measures or deflators, see \cite{Fernholz,Schal,Follmer-Kramkov} and the references therein.  \\

Recently in \cite{CL}, the authors consider the discrete-time model without transaction costs and addressed this super-hedging pricing issue differently and without any non-arbitrage assumption. As a result, the authors discovered that this infimum price is in fact a price if and only if  the model fulfills a weaker form of non-arbitrage, called {\it Absence of Immediate Profit} (AIP hereafter for short). This novel notion of non-arbitrage is weaker than the classical non-arbitrage concepts which all coincide in discrete-time setting.  Besides the AIP concept, using the conditional essential supremum as their main mathematical tool, the authors derived a backward equation/algorithm for calculating the super-hedging price process $\widehat{\cP}=(\widehat{\cP}_t)_{t=0,...,T}$ as follows
\begin{equation}\label{BSDEofCL}
\widehat{\cP}_t=\widehat{\cP}_{t,t+1}\Bigl(\widehat{\cP}_{t+1}\Bigr),\quad t=0,.., T-1,\quad\rm{and}\quad  \widehat{\cP}_T=C.
\end{equation} 
Here $\widehat{\cP}_{t,t+1}(\cdot)$ is the one-period super-hedging pricing operator for the period $t$, which can also be seen as the concave envelop of the payoff relatively to the convex envelop of the conditional support. The conditional essential supremum notion was introduced in \cite{BCJ}, \cite{KS} and developed in \cite{EL}, \cite{BCL}, \cite{CL}, \cite{KL} and the references therein to cite a few.


{\it What are our achievements?} In this informational setting, generated by the random horizon $\tau$, we describe explicitly the expansion of the set of super-hedging prices for various vulnerable claims. This expansion is quantified using processes under $\mathbb{F}$ and/or super-hedging prices of models under the flow $\mathbb{F}$. Besides, this shows how $\tau$ affect the valuation process, which is an important step towards addressing Immediate-Profit arbitrage for the stopped model $(S^{\tau},\mathbb{G})$. Precisely, we connect one-to-one the $\mathbb{G}$-Immediate-Profit arbitrage for the latter model to the Immediate-Profit arbitrage for  $({\cal{T}}_r(S),\mathbb{F}, \widehat{Q})$, where  ${\cal{T}}_r(S)$ is a transformed model of $S$  and  $\widehat{Q}$ is a probability measure quantifying the correlation risks generated by $\tau$ and $\mathbb{F}$. In this spirit, we show that the impact of $\tau$ on classical arbitrage differs tremendously from its impact on immediate-profit arbitrage. Our last achievement resides in determining the pricing formulas for vulnerable claims, in different manners. On the one hand, we show that for any vulnerable claim $H^{\mathbb{G}}$, there exists a unique pair $(f(t,\omega, x), H^{\mathbb{F}})$ such that the super-hedging price process of this claim coincide on $\Lbrack0,\tau\Lbrack$ with the solution of the following backward stochastic differential equations
\begin{equation}\label{BSDE4VO}
X_t=\widehat{\cP}_{t,t+1}\Bigl(f(t+1, X_{t+1})\Bigr),\quad X_T=H^{\mathbb{F}}).
\end{equation} 
Here $f(t,\omega, x)$ is an $\mathbb{F}$-adapted  functional intimately associated to the payment's policy of the claim and $\tau$, and it is not linear in $x$ in general. This extends the Carassus-Lepinette's pricing formula (\ref{BSDEofCL}) to more complex situation, and shows that (\ref{BSDEofCL}) remains valid for a subclass of vulnerable options only. On the other hand, we describe the dynamics of the super-hedging price process of the vulnerable claim and single out precisely the various induced informational risks. This latter decomposition formula is vital for risk management in the extended markets, in particular for the securitization of mortality and/or longevity risks in life insurance. All these aforementioned results rely essentially on understanding how conditional essential supremum behave under additional information and/or change of priors.

The paper has five sections including the current introductory section.  The second section presents the economical and mathematical model and gives some preliminaries. The third section addresses the essential supremum under change of probability and/or  change of filtration.  The fourth section discusses the super-hedging prices' set and the immediate-profit arbitrage, while the fifth section derives the pricing formulas for various vulnerable claims.
\section{The mathematical model and preliminaries}
Throughout the paper, we suppose given a complete probability space $(\Omega, {\cal{G}},P)$ and a fixed investment horizon $T\in(0,\infty)$. On this space, we suppose given the pair $(S, \mathbb{F})$. $\mathbb{F}:=(\cF_t)_{t=0,\cdots,T}$ is a filtration, which represents the flow of information available to all agents through time. $S=(S_t)_{t=0,...,T}$  is a $d$-dimensional and $\mathbb{F}$-adapted process with values in $\R^d_+=[0,\infty)^d$, and represents the (discounted) prices of $d$-risky assets.  Besides $S$, we suppose that there exists  a bond whose (discounted) price is $B_t=1$. Throughout the paper, the triplet $(S,\mathbb{F},P)$ will be called the {\it{initial model}}. Throughout the paper, $L^0(E,{\cal{H}})$ denotes the set of random variables having values in $E$ and are ${\cal{H}}$-measurable, for any set $E$ and any sub-$\sigma$-algebra $\cal{H}$. When $E=\RR$, we simply write $L^0({\cal{H}})$, while $L^0_+({\cal{H}})$ denotes its subset of nonnegative random variables.
\subsection{Random horizon and its parametrization}
Besides the initial model, we consider an arbitrary random time $\tau$, which might not be seen via $\mathbb{F}$ when  it occurs. Thus, in order to take into account the occurrence of this random time, we consider the progressive enlargement of $\mathbb{F}$ with $\tau$, and this yields to a larger flow $\mathbb{G}$ given by 
\bea \label{FiltrationG}\mathbb{G}:=({\cal{G}}_t)_{t=0,...,T},\ {\cal{G}}_t:={\cal{F}}_t\vee \sigma\left( \{\tau=r\}:\ r=0,\cdots,t\right),\ t=0,...,T.\eea 
The agent endowed with the flow $\mathbb{F}$ can see the occurrence of $\tau$ through the pair $(G,\widetilde{G})$ of two processes, called Az\'ema supermatingales, given by
\begin{equation}\label{ZandZtilde}
 G_t:=P\left(\tau>t\big|\ \cF_t\right),\quad \widetilde{G}_t:=P\left(\tau\ge t\big|\ \cF_t\right),\quad t=0,1,\cdots,T.
\end{equation}
The following lemma is a direct consequence of \cite[Lemma 4.6]{Jeulin}, see also \cite[(2.7)]{ChoulliDeng},  and is useful throughout the paper.
\begin{lemma}\label{Projection2G-F} For any $t\in\{1,...,T\}$, for any positive integer $n$, we have
$$L^0(\R^n,{\cal{G}}_{t-1})I_{\{\tau\geq t\}}=L^0(\R^n,{\cal{F}}_{t-1})I_{\{\tau\geq t\}}\ \mbox{and}\ L^0_+({\cal{G}}_{t-1})I_{\{\tau\geq t\}}=L^0_+({\cal{F}}_{t-1})I_{\{\tau\geq t\}}.$$
\end{lemma}
The following lemma characterizes inequalities in $\mathbb{G}$ using inequalities in $\mathbb{F}$. 
\begin{lemma}\label{IneqCharacterizationType2}  Let $t\in\{1,...,T\}$ and $(X_t,K_t)\in{L}^0(\cF_t)\times{L}^0(\cF_t)$. Then we have:\\
{\rm{(a)}} $X_t1_{\{\tau= t\}}\ge K_t1_{\{\tau= t\}}$ $P$-a.s. iff $X_t1_{\{\widetilde{G}_t>G_t\}}\ge K_t1_{\{\widetilde{G}_t>G_t\}}$ $P$-a.s..\\
{\rm{(b)}} $X_t1_{\{\tau> t\}}\ge K_t1_{\{\tau>t\}}$ $P$-a.s. iff $X_t1_{\{{G}_t>0\}}\ge K_t1_{\{G_t>0\}}$ $P$-a.s..\\
{\rm{(c)}} $X_t1_{\{\tau\ge t\}}\ge K_t1_{\{\tau\geq t\}}$ $P$-a.s. iff $X_t1_{\{\widetilde{G}_t>0\}}\ge K_t1_{\{\widetilde{G}_t>0\}}$ $P$-a.s..\\
{\rm{(d)}} $X_t1_{\{\tau\ge t\}}\ge K_t1_{\{\tau= t\}}$ $P$-a.s. iff $X_t1_{\{\widetilde{G}_t>0\}}\ge K_t1_{\{\widetilde{G}_t>G_t=0\}}+K_t^+1_{\{\widetilde{G}_t>G_t>0\}}$ $P$-a.s..
\end{lemma}
\begin{proof} It is clear that assertion (a) follows immediately due to  $(\tau=t)\subset(\widetilde{G}_t>G_t)$, while assertion (b) is due to $(\tau>{t})\subset(G_t>0)$. Also, $\{\tau\ge t\}\subseteq \{\widetilde{G}_t>0\}$ hence (c) holds. Thus the rest of the proof focuses on proving assertion (d). To this end, we notice that $X_tI_{\{\tau\ge t\}}\ge K_tI_{\{\tau= t\}}$ $P$-a.s. iff $X_tI_{\{\tau= t\}}\ge K_tI_{\{\tau= t\}}$ $P$-a.s. and $X_tI_{\{\tau> t\}}\ge 0$, $P$-a.s.. Thus, by combining these with assertions (a) and (b), we conclude that $X_tI_{\{\tau\ge t\}}\ge K_tI_{\{\tau= t\}}$ iff  $X_tI_{\{G_t>0\}}\geq 0$ $P$-a.s. and $X_tI_{\{\widetilde{G}_t>G_t\}}\geq K_tI_{\{\widetilde{G}_t>G_t\}}$ $P$.a.s.. Furthermore, it is easy to check that  $X_tI_{\{\widetilde{G}_t>G_t\}}\geq K_tI_{\{\widetilde{G}_t>G_t\}}$ and $X_tI_{\{G_t>0\}}\ge 0$ $P$-a.s. if and only if $X_t1_{\{\widetilde{G}_t>0\}}\ge K_t1_{\{\widetilde{G}_t>G_t=0\}}+K_t^+1_{\{\widetilde{G}_t>G_t>0\}}$ $P$-a.s.. This completes the proof of assertion (d), and ends the proof of the lemma.\end{proof}
\subsection{Super-hedging prices and Immediate-profit}
In this subsection, we consider an arbitrary market model $(X,\mathbb{H})$ defined on $(\Omega, {\cal{G}},P)$, where $\mathbb{H}$ is a filtration and $X$ is an $\mathbb{H}$-adapted process.\\
The following proposition states the existence of the conditional supremum and infimum of any family of random variables, as shown in \cite[Section 5.3.1]{KS}.
\begin{proposition}Let $\cH\subseteq \cF$ be two sub $\sigma$-algebras.  For any family of random variables $\Gamma\subseteq L^0([-\infty,\infty],\cF)$, there exists a unique (up to a negligible set) $\cH$-measurable random variable $\gamma_{\cH}\in L^0([-\infty,\infty],\cH)$ such that $\gamma_{\cH}\ge \gamma$, for all $\gamma\in \Gamma$, and if $\gamma_{\cH}^1\in L^0([-\infty,\infty],\cH)$ is such that $\gamma_{\cH}^1\ge \gamma$ for all $\gamma\in \Gamma$, then $\gamma_{\cH}^1\ge \gamma_{\cH}$ a.s..We call $\gamma_{\cH}$ the conditional supremum of  $\Gamma$ knowing $\cH$ and we denote it by $\underset{{\cH}}{\esssup}(\Gamma)$. Similarly, we define $\underset{{\cH}}{\essinf}(\Gamma)=-\underset{{\cH}}{\esssup}\left(-\Gamma\right)$.
\end{proposition}
When $\cal{H}=\cal{G}$ in the proposition above, we write $\underset{\cal{H}}{\esssup}(\Gamma)=\esssup(\Gamma)$.   Let $\cP^a(\bP)$ be the set of all absolutely continuous probability measures w.r.t. $\bP$. 

\begin{lemma}\label{TowerProperty4Essup}Let ${\cal{H}}_1\subseteq {\cal{H}}_2\subseteq {\cal{G}}$ be sub-$\sigma$-algebras and let $\Gamma\subseteq L^0([-\infty,\infty], {\cal{G}})$ be a family of random variables. Then the following properties hold.
\begin{equation}\label{eualities4TowerProperty}
\underset{{\cal{H}}_2}\esssup(\Gamma)\leq  \underset{{\cal{H}}_1}\esssup(\Gamma),\quad \mbox{and}\quad \underset{{\cal{H}}_1}\esssup(\Gamma)=  \underset{{\cal{H}}_1}\esssup\left( \underset{{\cal{H}}_2}\esssup(\Gamma)\right).
\end{equation}
\end{lemma}
\begin{lemma} \label{f(esssup)} If $\cH\subseteq \cG$ is a sub-$\sigma$-algebra and $\Gamma \subseteq L^0((-\infty,\infty],\cG)$, then 
\begin{equation}\label{essential4+}
\left( \underset{{\cH}}\esssup(\Gamma )\right)^+= \underset{{\cH}}\esssup(\Gamma ^+),\quad\mbox{and}\quad \left( \underset{{\cH}}\esssup(\Gamma )\right)^-
= \underset{{\cH}}\essinf(\Gamma ^-)\end{equation}
\end{lemma}
\begin{proof}
Note that the first equality in (\ref{essential4+}) can easily be shown, and hence its proof will be omitted. From $ \underset{{\cH}}\esssup(\Gamma )\ge Y$ $P$-a.s. for any $Y\in \Gamma$, we get  
$\left( \underset{{\cH}}\esssup(\Gamma )\right)^-\le Y^-$ $P$-a.s., and hence $\left( \underset{{\cH}}\esssup(\Gamma )\right)^-\leq \underset{{\cH}}\essinf(\Gamma ^-)$. Then remark that $ \underset{{\cH}}\essinf(\Gamma ^-)\le Y^-$ yields $Y^+-\essinf_{\cH}(\Gamma ^-)\ge Y^+-Y^-$. Thus, $ \underset{{\cH}}\esssup(\Gamma ^+)- \underset{{\cH}}\essinf(\Gamma ^-)\ge Y$, and $ \underset{{\cH}}\esssup(\Gamma ^+)- \underset{{\cH}}\essinf(\Gamma ^-)\geq \underset{{\cH}}\esssup(\Gamma )$. By the first equality in (\ref{essential4+}), $ \underset{{\cH}}\esssup(\Gamma ^+)=\left( \underset{{\cH}}\esssup(\Gamma )\right)^+$ so we deduce that $- \underset{{\cH}}\essinf(\Gamma ^-)\geq -\left( \underset{{\cH}}\esssup(\Gamma )\right)^-$ and, finally $ \underset{{\cH}}\essinf(\Gamma ^-)\le \left( \underset{{\cH}}\esssup(\Gamma )\right)^-$. Then the second equality in (\ref{essential4+}) follows, and the proof is complete.\end{proof}
This lemma can be extended easily to increasing and decreasing functions. \\
Recall that a self-financing portfolio process in discrete-time is a stochastic process $(V_t)_{t=0}^T$ satisfying $\Delta V_t:=V_t-V_{t-1}=\theta_{t-1}\Delta{X}_t$ for some $\theta_{t-1}\in L^0(\R^d,{\cal{H}}_{t-1})$. We denote by ${\cal{A}}_{t,u}:={\cal{R}}_{t,u}-L^0_+({\cal{H}}_u)$, $t\le u \le T$, the set of all attainable claims at time $u$ when starting a self-financing portfolio process from the zero initial endowment at time $t$. By definition, $V_{t,u}\in  {\cal{R}}_{t,u}$ if and only if $V_{t,u}=\sum_{r=t+1}^u\theta_{r-1}\Delta S_r$ for some $\theta_{r}\in L^0(\R^d,{\cal{H}}_r)$, $r=t,\cdots,u-1$. \\
For any payoff $\xi\in L^0({\cal{H}}_T)$, we associate the pair $({\cal{P}}_{t}(\xi), \Pi^*_t(\xi))$ of the set of all super-hedging prices and the infimum price, for $t=0,1,...,T-1,$  given by
\begin{equation}
\begin{split}
{\cal{P}}_{t}(\xi)&:=\left\{p_t\in{L}^0(\R,{\cal{H}}_t):\ p_t+V_{t,T}\geq \xi,\ P\mbox{-a.s.}\ \mbox{for}\  V_{t,T}\in {\cal{R}}_{t,T}\right\}, \\
{\widehat{\cal{P}}}_t(\xi)&:=\essinf ({\cal{P}}_{t}(\xi)),\quad t=0,1,...,T-1.
\end{split}
\end{equation}
Thus, it is easy to check that  ${\cal{P}}_{t}(0)=(-{\cal{A}}_{t,T}) \cap L^0({\cal{H}}_t)$. 
\begin{definition}\label{Definition4MinimumPrice} For a model $(X, \mathbb{H}:=(\cH_t)_{t=0}^T)$, a probability $Q$, two dates $t_1,t_2$ such that $0\leq t_1<t_2$, and a payoff $\xi\in L^0({\cal{H}}_{t_2})$, we denote by ${\cal{P}}_{t_1,t_2}^{(X,\mathbb{H},Q)}(\xi)$ the set of all super-hedging prices at time $t_1$ of the payoff $\xi$. The infimum price of $\xi$, denoted by $\widehat{\cal{P}}_{t_1,t_2}^{(X,\mathbb{H},Q)}(\xi)$ is given by 
 \begin{equation}\label{infinimumPrice}
 \widehat{\cal{P}}_{t_1,t_2}^{(X,\mathbb{H},Q)}(\xi):=\essinf {\cal{P}}_{t_1,t_2}^{(X,\mathbb{H},Q)}(\xi).
 \end{equation}
 When $Q=P$, we omit the probability in the notation, and write  ${\cal{P}}_{t_1,t_2}^{(X,\mathbb{H})}(\xi)$ and $\widehat{\cal{P}}_{t_1,t_2}^{(X,\mathbb{H})}(\xi)$ instead.

\end{definition}
 Below, we recall, from \cite{CL}, the mathematical definition of the AIP concept.
\begin{definition} We say that the condition AIP holds at time $t\leq T-1$ if ${\cal{A}}_{t,T}\cap L^0_+({\cal{F}}_t)=\{0\}$. We say that AIP holds if it holds at any time $t\leq T-1$.
\end{definition}
 Similarly, as for the classical non-arbitrage condition in discrete-time, the AIP concept can be checked step-by-step and in various manners. This is the aim of the following proposition.
\begin{proposition}\label{GeneralAIP} Let $\mathbb{H}=({\cal{H}}_t)_{t=0,..,T}$ be a filtration, and $X$ be an $\mathbb{H}$-adapted process. Then the following assertions are equivalent.\\
{\rm{(a)}} The model $(X, \mathbb{H})$ satisfies AIP.\\
{\rm{(b)}}  For $t\in\{1,...,T\}$ and $\theta\in L^0(\R^d,{\cal{H}}_{t-1})$,  $\underset{\cH_{t-1}}\esssup( \theta\Delta{X}_t)\ge 0$ $P$-a.s..\\
{\rm{(c)}}  For any $t\in\{1,...,T\}$, ${\cal{P}}^{(X,\mathbb{H})}_{t-1,t}(0)\subseteq{L}^0_+({\cal{H}}_{t-1})$. \\
{\rm{(d)}}  For any $t\in\{1,...,T\}$, $\widehat{\cal{P}}^{(X,\mathbb{H})}_{t-1,t}(0)=0$, $P$-a.s..
\end{proposition}
\begin{proof}
By \cite[Proposition 2.11]{CL}, a model $(X, (\cH_t)_{t=0}^T)$ satisfies AIP at time $t-1\ge 0$ if and only if,  $0$ belongs to the closed convex hull of $\bar D_{t-1}={\rm supp}_{\cH_{t-1}}(\Delta X_t)$, the random conditional support of $\Delta X_t$. By the Hahn-Banach separation theorem for convex sets, this is equivalent to the property that  $\sigma_{\bar D_{t-1}}(x)\ge 0$, for all $x\in \R^d$, a.s.($\omega$), where $\sigma_{\bar D_{t-1}}(x)=\sup_{z\in \bar D_{t-1}}(-xz)$. Note that, by \cite[Theorem 3.4]{EL}, $\sigma_{\bar D_{t-1}}(x)=\underset{\cH_{t-1}}\esssup(-x\Delta X_t)$. Therefore, a model $(X, (\cH_t)_{t=0}^T)$ satisfies AIP at time $t-1\ge 0$ iff $\underset{\cH_{t-1}}\esssup(-x\Delta X_t)\ge 0$ for $x\in \R^d$, a.s.. Hence, we get $\underset{\cH_{t-1}}\esssup( \theta_{t-1}\Delta X_t)\geq 0$ for $\theta_{t-1}\in L^0(\R^d,\cH_{t-1})$. Indeed, the inequality  $\underset{\cH_{t-1}}\esssup( \theta_{t-1}\Delta X_t)\ge 0$ is immediate by the previous reasoning if AIP holds. Reciprocally, if $ \underset{\cH_{t-1}}\esssup(\theta_{t-1}\Delta X_t)\geq 0$ for $ \theta_{t-1}\in L^0(\R^d,\cH_{t-1})$, then $\underset{\cH_{t-1}}\esssup(-x\Delta X_t)\ge 0$ for all $x\in \R^d$, a.s.. To see it, we argue by contradiction and, by a measurable selection, it is possible to get $\underset{\cH_{t-1}}\esssup(\theta_{t-1}\Delta X_t)< 0$  on a non null set, for some $ \theta_{t-1}\in L^0(\R^d,\cH_{t-1})$, i.e. a contradiction. This ends the proof of the proposition.\end{proof}
\section{Essential supremum under change of priors or information}
Herein, we derive novel properties on conditional essential supremum, which are very useful throughout the paper. In  fact, we investigate how the conditional essential supremum behaves under a change of probability and change of information (filtration). To this end, we start with the easy but crucial lemma, which conveys that  conditional essential supremum or infinimum of indicators is always an indicator. 
\begin{lemma}\label{CharEssIndBis}Let $\cH_1\subseteq \cH$ be two sub-$\sigma$-algebras, and $H\in \cH$. Then the following assertions hold.\\
{\rm{(a)}} There exists $H_1\in \cH_1$ satisfying
\begin{equation}\label{EssInf2Indicator1Bis}
\underset{\cH_1}{\essinf}(I_H)=I_{H_1}=\displaystyle{I}_{\{\underset{\cH_1}{\essinf}(1_H)>0\}}=\displaystyle{I}_{\{\underset{\cH_1}{\essinf}(1_H)=1\}},\quad P\mbox{-a.s.,}\end{equation}
 and it is the largest  $\cH_1$-measurable set contained in $H$.\\
{\rm{(b)}} There exists $H_2\in \cH_1$ such that  
\begin{equation}\label{EssInf2Indicator2Bis}
\underset{\cH_1}{\esssup}(I_H)=I_{H_2}=\displaystyle{I}_{\{ \underset{\cH_1}{\esssup}(1_H)=1\}}=\displaystyle{I}_{\{\underset{\cH_1}{\esssup}(1_H)>0\}},\quad P\mbox{-a.s.,}\end{equation}
 and it is the smallest  $\cH_1$-measurable set containing $H$.
\end{lemma}
\begin{proof}
Remark that assertion  (a) follows immediately from  assertion (b) applied  to $\overline{H}:=\Omega\setminus{H}\in{\cal{H}}$, and the easy fact that $\underset{\cH_1}{\esssup}(I_H)=1- \underset{\cH_1}{\essinf}(I_{\overline{H}})$. Thus, the rest of this proof focuses on assertion (b). To this end, we put 
$$H_2:=\Bigl\{\underset{\cH_1}{\esssup}(I_{H})=1\Bigr\},$$
which belongs to ${\cal{H}}_1$. In  virtue of  the definition of  conditional essential supremum, we always have $\esssup_{\cH_1}(I_{H})\geq{I}_H$ $P$-a.s., and hence $I_H\leq{I}_{H_2}$, $P$-a.s.. Therefore, we get 
$$\underset{\cH_1}{\esssup}(I_H)\leq {I}_{H_2},\quad P\mbox{-a.s.}.$$
As a result of this, we obtain 
$$\{\underset{\cH_1}{\esssup}(I_H)>0\}=\{\underset{\cH_1}{\esssup}(I_H)=1\}=H_2,\quad P\mbox{-a.s.},$$
and (\ref{EssInf2Indicator2Bis}) follows immediately. This ends the proof of the lemma.\end{proof}
\begin{theorem}\label{ChangeofProbaFiltration} Let ${\cal{H}}_i$, $i=1,2$ be two sub-$\sigma$-algebras of $\cal{G}$ such that ${\cal{H}}_1\subseteq{\cal{H}}_2$, and  $Z\in L_+({\cal{H}}_2)$ with $E[Z]=1$, and put
$$Z^{{\cal{H}}_1}:=E[Z\ \big|{\cal{H}}_1]\quad\mbox{and}\quad Q:=Z\cdot{P}.$$
Consider $\Gamma\subset{L}^{0}({\cal{H}}_2)$, and denote by 
\begin{equation}\label{gammatilde(Q,P)}
\widetilde{\gamma}^Q:= \underset{{\cal{H}}_1}{\overset{Q} {\esssup}}(\Gamma)\quad\mbox{and}\quad \widetilde{\gamma}:=\underset{{\cal{H}}_1}{\esssup}(\Gamma{I}_{\{Z>0\}}).\end{equation}
 Then the following assertions hold.\\
{\rm{(a)}} If $X\in L_+^0({\cal{H}}_2)$ and $Y:=E\left[X\ \Big|\ {\cal{H}}_1\right]$, then $P$-a.s. we have 
\begin{equation}\label{EssentialSup(H1,H2)}
\begin{split}
&\underset{{\cal{H}}_1} {\esssup}(I_{\{X>0\}})=I_{\{Y>0\}},\ \rm{and}\  \underset{{\cal{H}}_1} {\essinf}(I_{\{X>0\}})=I_{\{P(X>0|{\cal{H}}_1)=1\}},\\
&\underset{{\cal{H}}_1} {\esssup}(I_{\{X=0\}})=I_{\{P(X=0|{\cal{H}}_1)>0\}},\ \rm{and}\  \underset{{\cal{H}}_1} {\essinf}(I_{\{X=0\}})=I_{\{Y=0\}}.\end{split}
\end{equation}
{\rm{(b)}} We always have 
\begin{equation}\label{Positivity4essentialSup}
\begin{split}
\widetilde{\gamma}\geq\widetilde{\gamma}^Q\quad{Q}\mbox{-a.s.}\quad\mbox{and}\quad  {I}_{\{P(Z=0|{\cal{H}}_1)>0\}}\widetilde{\gamma}\geq0\quad P\mbox{-a.s.}.\end{split}
 \end{equation}
{\rm{(c)}} We have $\widetilde{\gamma}^Q=\widetilde{\gamma}$, $Q$-a.s. on $(\widetilde{\gamma}^Q\geq 0)$ ( i.e. $\widetilde{\gamma}^QI_{\{\widetilde\gamma^Q\geq 0\}}=\widetilde{\gamma}I_{\{\widetilde\gamma^Q\geq 0\}}$ $Q$-a.s.).\\
{\rm{(d)}} If we denote  $\Gamma^+:=\left\{\gamma^+:\ \gamma\in\Gamma\right\}$, then we get
\begin{equation}\label{EssentialSup(H1,Q)Bis}
\left(\widetilde{\gamma}^Q\right)^+=\underset{{\cal{H}}_1}{\overset{Q} {\esssup}}(\Gamma^+)=(\widetilde{\gamma})^+,\quad Q\mbox{-a.s..}\end{equation}
 {\rm{(e)}} On $\left(P(Z=0|\ {\cal{H}}_1)=0\right)$, we have 
  \begin{equation*}\widetilde{\gamma}^Q=\widetilde{\gamma}=\underset{{\cal{H}}_1}{\esssup}(\Gamma)\quad P\mbox{-a.s}..\end{equation*}
 {\rm{(f)}} If  $\widetilde{\gamma}^Q\geq 0$ $Q$-a.s., then $\widetilde{\gamma}\geq 0$ $P$-a.s.. Furthermore, we have 
 \begin{equation}\label{EssentialSup-Signs}
  \left(\widetilde{\gamma}^Q> 0\right)=\left(\widetilde{\gamma}> 0\right)\quad Q\mbox{-a.s.}\quad \mbox{and}\quad \underset{{\cal{H}}_1}{\esssup}(\Gamma{I}_{\{Z=0\}})\geq 0\quad Q\mbox{-a.s.}.\end{equation}
 In particular, $\widetilde{\gamma}^Q> 0$ $Q$-a.s.  if and only if $\widetilde{\gamma}> 0$ $Q$-a.s..\\
    {\rm{(g)}}  For $P$-almost surely, we have 
 \begin{equation}\label{EssentialSup-SignsunderP}\begin{split}
& \left(\widetilde{\gamma}^Q<\widetilde{\gamma}\right)=\left(\widetilde{\gamma}^Q<0\right)\cap\left(P(Z=0|\ {\cal{H}}_1)>0\right)=\left(\widetilde{\gamma}^Q<0\leq \widetilde{\gamma}\right), \end{split}
 \end{equation}
and 
 \begin{equation}\label{EssentialSup-SignsunderP1} \left(\widetilde{\gamma}<0\right)=\Bigl(P(Z=0\big|{\cal{H}}_1)=0\Bigr)\cap\left(\widetilde{\gamma}^Q<0\right).
 \end{equation}
\end{theorem}
It worths mentioning that similar results for essential infimum can be obtained easily. In fact, thanks to the fact that $\underset{{\cal{H}}_1}{\essinf}(-\Gamma)=-\underset{{\cal{H}}_1}{\esssup}(\Gamma)$ for any $\Gamma\subset{L}^0({\cal{H}}_2)$, the following corollary can be easily proved.
\begin{corollary} Consider the notations of Theorem \ref{ChangeofProbaFiltration}. Then the following assertions hold.\\
{\rm{(a)}} Suppose $\Gamma\subset{L}^{0}({\cal{H}}_2,P)$. Then we have 
\begin{equation}\label{Negativity4essentialSup}
\begin{split}
& \underset{{\cal{H}}_1}\essinf(\Gamma{I}_{\{Z>0\}})\leq \underset{{\cal{H}}_1}{\overset{Q} {\essinf}}(\Gamma),\quad P\mbox{-a.s.}\quad\mbox{on}\quad \Bigl(Z^{{\cal{H}}_1}>0\Bigr),\\
& \underset{{\cal{H}}_1}\essinf(\Gamma{I}_{\{Z>0\}})\leq0\quad P\mbox{-a.s.}\quad\mbox{on}\quad \Bigl(P\left(Z=0\big|{\cal{H}}_1\right)>0\Bigr).\end{split}
 \end{equation}
{\rm{(b)}}  If $\Gamma\subset{L}^{0}({\cal{H}}_2,P)$ and $ \underset{{\cal{H}}_1}{\overset{Q} {\essinf}}(\Gamma)\leq 0$ $Q$-a.s., then 
\begin{equation}\label{EssentialSup(H1,Q)}
\underset{{\cal{H}}_1}{\overset{Q} {\essinf}}(\Gamma)= \underset{{\cal{H}}_1} {\essinf}(\Gamma{I}_{\{Z>0\}})\quad P\mbox{-a.s. on}\quad (Z^{{\cal{H}}_1}>0).
\end{equation}
{\rm{(c)}  If $ \underset{{\cal{H}}_1}{\overset{Q} {\essinf}}(\Gamma)\leq 0$ $Q$-a.s.}, then $ \underset{{\cal{H}}_1} {\essinf}(\Gamma{I}_{\{Z>0\}})\leq 0$ $P$-a.s..

\end{corollary}
The last two assertions in Theorem \ref{ChangeofProbaFiltration} clearly describes how the sign of $\widetilde\gamma^Q$ is related to the sign of $\widetilde\gamma$. This fact is very important, as the sign of conditional essential supremum plays a crucial role in arbitrage. In fact, the equality (\ref{EssentialSup-SignsunderP}) conveys that the two essential supremums, $\widetilde\gamma^Q$ and $\widetilde\gamma$, have different signs if and only if they do not coincide. However, the two essential supremums coincide when they have same sign, see (\ref{EssentialSup-SignsunderP1}) and assertion (c).\\
The converse of the first statement in assertion (f) is not true in general. In the following, we illustrate an example to support this claim.
\begin{example} Consider $Q$ such that $P(Z=0<Z^{{\cal{H}}_1})>0$, and for some $\epsilon\in(0,\infty)$ put $\Gamma:=\left\{-\epsilon{I}_{\{P(Z=0\ |{\cal{H}}_1)>0\}}\right\}$. Then direct calculations yield 
\begin{equation*}
\begin{split}
\underset{{\cal{H}}_1}{\esssup}(\Gamma{I}_{\{Z>0\}})&=-\epsilon{I}_{\{P(Z=0\ |{\cal{H}}_1)>0\}}\underset{{\cal{H}}_1}{\essinf}(I_{\{Z>0\}})\\
&=-\epsilon{I}_{\{P(Z=0\big|{\cal{H}}_1)>0\}}{I}_{\{P(Z>0\ |{\cal{H}}_1)=1\}}=0,\quad P\rm{-a.s.},\\
\underset{{\cal{H}}_1}{\overset{Q} {\esssup}}(\Gamma)&=-\epsilon{I}_{\{P(Z=0\ |{\cal{H}}_1)>0\}},\end{split}\end{equation*}
and $Q(P(Z=0\ |{\cal{H}}_1)>0)=E[Z^{{\cal{H}}_1}1_{\{P(Z=0\ |{\cal{H}}_1)>0\}}]>0$ by assumption. Therefore,  for this choice of $\Gamma$, these latter statements prove that we have $\underset{{\cal{H}}_1}{\esssup}(\Gamma{I}_{\{Z>0\}})\geq 0$ $P$-a.s. and $Q\left(\underset{{\cal{H}}_1}{\overset{Q} {\esssup}}(\Gamma)<0\right)>0$. Similar conclusion holds for $\Gamma:=\left\{-\epsilon{I}_{\{P(Z=0\ |{\cal{H}}_1)>0\}}+\epsilon{I}_{\{P(Z=0\ |{\cal{H}}_1)=0\}}\right\}$.
\end{example}
\begin{proof}{\it of Theorem \ref{ChangeofProbaFiltration}.} On the one hand, by combining assertions (f) and (d), we deduce that $\widetilde\gamma^Q=\widetilde\gamma$ $Q$-a.s. as soon as $\widetilde\gamma^Q\geq 0$ $Q$-a.s.. On the other hand, in order to prove assertion (c), we apply the latter claim to $\Gamma':=\Gamma{I}_{\{\widetilde\gamma^Q\geq 0\}}$, which satisfies $\underset{{\cal{H}}_1}{\overset{Q} {\esssup}}(\Gamma')=(\widetilde\gamma^Q)^+\geq 0$ $Q$-a.s., and get 
$$
(\widetilde\gamma^Q)^+=\underset{{\cal{H}}_1}{\overset{Q} {\esssup}}(\Gamma')=\underset{{\cal{H}}_1}{\esssup}(\Gamma'{I}_{\{Z>0\}})=\widetilde\gamma{I}_{\{\widetilde\gamma^Q\geq 0\}},\quad Q\mbox{-a.s.}.$$
 This ends the proof of assertion (c), and the rest of this proof focuses on assertions (a), (b), (d), (e), (f) and (g) in three parts.\\
{\bf Part 1.} Hereto we prove assertion (a).  Let $X\in L_+^0({\cal{H}}_2)$, put $X_n:=\min(n,X)$ and $Y_n:=E[X_n\big|{\cal{H}}_1]$, and remark that both $X_n$ and $Y_n$ increase to $X$ and $Y$ respectively, and 
$$
(X_n>0)=(X>0)\quad\mbox{and}\quad  (Y_n>0)=(Y>0),\quad n\geq 1.$$
Thus, it is enough to prove the assertion for bounded $X\in L_+^0({\cal{H}}_2)$, and without loss of generality we assume that $\Vert{X}\Vert_{\infty}= 1$. On the one hand, it is clear that we always have $(Y=0)\subset(X=0)$, or equivalently  $I_{\{X>0\}}\leq{I}_{\{Y>0\}}$ $P$-a.s., which yields 
\begin{equation}\label{X2Y}
\underset{{\cal{H}}_1}{\esssup}(I_{\{X>0\}})\leq{I}_{\{Y>0\}},\quad P\mbox{-a.s.}.\end{equation}
On the other hand, we have $X=XI_{\{X>0\}}\leq I_{\{X>0\}}\leq\underset{{\cal{H}}_1} {\esssup}(I_{\{X>0\}})$ $P$-a.s., and hence by taking conditional expectation, we get $Y\leq\underset{{\cal{H}}_1}{\esssup}(I_{\{X>0\}})$, $P$-a.s.. This clearly implies that $\{Y>0\}\subset\{\underset{{\cal{H}}_1}{\esssup}(I_{\{X>0\}})>0\}$, and by combining this with Lemma \ref{CharEssIndBis}-(b), we deduce that
$$I_{\{Y>0\}}\leq I_{\{\underset{{\cal{H}}_1}{\esssup}(I_{\{X>0\}})>0\}}=\underset{{\cal{H}}_1}{\esssup}(I_{\{X>0\}}).$$ Therefore, by combining this latter inequality with (\ref{X2Y}), the first equality in (\ref{EssentialSup(H1,H2)}) follows immediately, while the fourth equality is a direct consequence of the first equality and 
$$
\underset{{\cal{H}}_1}{\essinf}(I_{\{X=0\}})=1-\underset{{\cal{H}}_1}{\esssup}(I_{\{X>0\}}).$$
Similarly, remark that the second equality in (\ref{EssentialSup(H1,H2)}) is a direct consequence of the third equality. To prove the third equality, we remark that  
$$I_{\{X=0\}}\leq I_{\{\underset{{\cal{H}}_1} {\esssup}(I_{\{X=0\}})>0\}}=\underset{{\cal{H}}_1}{\esssup}(I_{\{X=0\}})\quad P\rm{-a.s.}.$$
Then by taking conditional expectations on both sides, we get 
$$ P(X=0|{\cal{H}}_1)\leq I_{\{\underset{{\cal{H}}_1}{\esssup}(I_{\{X=0\}})>0\}}.$$ On the one hand, this clearly yields
\begin{equation}\label{firstincludion}
( P(X=0|{\cal{H}}_1)>0) \subseteq\left(\underset{{\cal{H}}_1}{\esssup}(I_{\{X=0\}})>0\right).\end{equation}
On the other hand, due to the easy fact that $(X=0)\subset( P(X=0|{\cal{H}}_1)>0)$, which is equivalent to $I_{\{X=0\}}\leq I_{\{ P(X=0|{\cal{H}}_1)>0\}}$ $P$-a.s., we get 
$$\underset{{\cal{H}}_1}{\esssup}(I_{\{X=0\}})\leq  I_{\{P(X=0|{\cal{H}}_1)>0\}}\quad P\rm{-a.s..}$$
Thus, by combining this last inequality with (\ref{firstincludion}) and again Lemma \ref{CharEssIndBis}, the third equality in (\ref{EssentialSup(H1,H2)}) follows immediately. This proves assertion (a).\\
{\bf Part 2.} This part proves assertions (b) and (d).  To this end, we consider $\Gamma\subset{L}^0({\cal{H}}_2)$, $Q\ll{P}$ with density $Z:=dQ/dP$, and $Z^{{\cal{H}}_1}:=E[Z\big|{\cal{H}}_1]$. Then it is clear that 
$$
\underset{{\cal{H}}_1}{\esssup}(\Gamma{I}_{\{Z>0\}})\geq \gamma{I}_{\{Z>0\}}\quad P\mbox{-a.s.},\quad \forall\ \gamma\in\Gamma.$$
This is equivalent to  
\begin{equation}\label{Equivalence10}
\begin{split}
\widetilde{\gamma}\geq \gamma\quad Q\mbox{-a.s.},\quad  \mbox{for any}\ \gamma\in\Gamma,\quad \rm{and}\quad {I}_{\{Z=0\}}\widetilde{\gamma}\geq 0\quad P\mbox{-a.s.}
\end{split}
\end{equation}
Hence, on the one hand, the first inequality above yields  the first inequality in (\ref{Positivity4essentialSup}) (i.e. $
\widetilde{\gamma}\geq\widetilde{\gamma}^{Q}$ $Q$-a.s.).
 On the other hand, the second inequality in (\ref{Equivalence10}) is equivalent to 
 $
 I_{\{Z=0\}}\leq{I}_{\{\widetilde{\gamma}\geq 0\}}$  $P$-a.s.. Then  by taking conditional essential supremum and using assertion (a), the second inequality in (\ref{Positivity4essentialSup}) follows. This ends the proof of assertion (b). To prove assertion (d), on the one hand, we notice that the first equality in (\ref{EssentialSup(H1,Q)Bis}) is due to Lemma \ref{f(esssup)}. On the other hand, we suppose that $\widetilde{\gamma}\geq 0$ $P$-a.s. and thanks to assertion (b) (see (\ref{Positivity4essentialSup})) we get 
\begin{equation}\label{FirstInequalityBis}
I_{\{Z^{{\cal{H}}_1}>0\}}\widetilde{\gamma}\geq (\widetilde{\gamma}^{Q})^+{I}_{\{Z^{{\cal{H}}_1}>0\}},\quad P\mbox{-a.s.}.\end{equation}
Furthermore, thanks to assertion (a), the definition of essential supremum, and Lemma \ref{f(esssup)}, we get  for any $\gamma\in\Gamma$,
$$
{I}_{\{Z^{{\cal{H}}_1}>0\}}(\widetilde{\gamma}^{Q})^+=\underset{{\cal{H}}_1}{\esssup}\left(I_{\{Z>0\}}(\widetilde{\gamma}^{Q})^+\right)\geq\gamma^+{I}_{\{Z>0\}}\geq \gamma{I}_{\{Z>0\}}\quad P\mbox{-a.s.}.$$
Hence, by taking conditional essential supremum, we obtain
$$
I_{\{Z^{{\cal{H}}_1}>0\}}(\widetilde{\gamma}^{Q})^+\geq\underset{{\cal{H}}_1} {\esssup}(\Gamma{I}_{\{Z>0\}})\geq {I}_{\{Z^{{\cal{H}}_1}>0\}}\widetilde{\gamma},\quad P\mbox{-a.s.}$$
Therefore, by combining this latter inequality with (\ref{FirstInequalityBis}), (\ref{EssentialSup(H1,Q)Bis}) holds under the assumption $\widetilde\gamma\geq0$ $P$-a.s.. To prove (\ref{EssentialSup(H1,Q)Bis}) in general, we put $\Gamma':=\Gamma{I}_{\{\widetilde\gamma\geq0\}}$ and derive 
$$\underset{{\cal{H}}_1}{\esssup}(\Gamma'{I}_{\{Z>0\}})=(\widetilde\gamma)^+\geq 0\quad\mbox{and}\quad 
\underset{{\cal{H}}_1}{\overset{Q} {\esssup}}(\Gamma')={\widetilde\gamma}^QI_{\{\widetilde\gamma\geq0\}}.$$
Thus, in particular we have $\underset{{\cal{H}}_1}{\esssup}(\Gamma'{I}_{\{Z>0\}})\geq 0$ $P$-a.s., and hence we can apply  (\ref{EssentialSup(H1,Q)Bis}) to $\Gamma'$ and derive
$$ ({\widetilde\gamma}^Q)^+=({\widetilde\gamma}^Q)^+I_{\{\widetilde\gamma\geq0\}}=(\underset{{\cal{H}}_1}{\overset{Q} {\esssup}}(\Gamma'))^+=\underset{{\cal{H}}_1}{\esssup}(\Gamma')=(\widetilde\gamma)^+,\quad Q\mbox{-a.s.}.$$
The first equality follows from $(\widetilde\gamma^Q\geq 0)\subset(\widetilde\gamma\geq 0)$ $Q$-a.s., which is due to assertion (b). This proves assertion (d), and ends part 2.\\ 
{\bf Part 3.} Herein, we prove assertions (e), (f) and (g).  For assertion (e), we use $\Sigma:=(P(Z=0|{\cal{H}}_1)=0)\subset(Z>0)\subset(Z^{{\cal{H}}_1}>0)$, and obtain  $\widetilde\gamma{I}_{\Sigma}=\underset{{\cal{H}}_1}{\esssup}(\Gamma{I}_{\Sigma\cap\{Z=0\}})=\underset{{\cal{H}}_1}{\esssup}(\Gamma)$ and $I_{\Sigma}\underset{{\cal{H}}_1}{\overset{Q} {\esssup}}(\Gamma)=I_{\Sigma}\underset{{\cal{H}}_1}{\esssup}(\Gamma)$. Thus, assertion (e)  follows immediately from these equalities.\\
To prove assertion (f), we start by noticing that due to assertion (b), the inequality $\widetilde{\gamma}^Q\geq 0$ $Q$-a.s. implies that $\widetilde\gamma\geq 0$ $Q$-a.s., and this is equivalent to $\widetilde\gamma\geq 0$ $P$-a.s. on $(Z^{{\cal{H}}_1} >0)$. Furthermore, thanks to $(Z^{{\cal{H}}_1} =0)\subset(Z=0)$ $P$-a.s., it is clear that $\widetilde\gamma=0$ $P$-a.s. on $(Z^{{\cal{H}}_1} =0)$. Thus, we get $\widetilde\gamma\geq 0$ $P$-a.s., and the first claim in assertion (f) is proved. \\
It is obvious that the first property in (\ref{EssentialSup-Signs}) follows immediately from (\ref{EssentialSup(H1,Q)Bis}). To prove the second property in (\ref{EssentialSup-Signs}), we use $\underset{{\cal{H}}_1}{\esssup}(\Gamma{I}_{\{Z=0\}})\geq \gamma{I}_{\{Z=0\}}$ $P$-a.s. for any $\gamma\in\Gamma$, and hence we get $\underset{{\cal{H}}_1}{\esssup}(\Gamma{I}_{\{Z=0\}})\geq 0$ $Q$-a.e.. This proves the third claim in assertion (f), and completes the proof of assertion (f). \\
To prove assertion (g), we derive $(\widetilde\gamma<0)\subset (P(Z=0|{\cal{H}}_1)=0)\cap(\widetilde\gamma^Q<0)$ due to assertion (b), while $(P(Z=0|{\cal{H}}_1)=0)\cap(\widetilde\gamma^Q<0)\subset (\widetilde\gamma<0)$ follows immediately from assertion (e). This proves (\ref{EssentialSup-SignsunderP1}).\\
Remark that $(\widetilde\gamma^Q<\widetilde\gamma)\subset (\widetilde\gamma^Q<0)$, in virtue of assertion (c), and due to assertion (e) we have  $(\widetilde\gamma^Q<\widetilde\gamma)\subset (P(Z=0|{\cal{H}}_1)>0)$. Thus, by combining these latter two remarks and the second property of assertion (b), we obtain
$$(\widetilde\gamma^Q<\widetilde\gamma)\subset (\widetilde\gamma^Q<0)\cap (P(Z=0|{\cal{H}}_1)>0)\subset (\widetilde\gamma^Q<0)\cap(\widetilde\gamma\geq0)\subset (\widetilde\gamma^Q<\widetilde\gamma).$$
This proves (\ref{EssentialSup-SignsunderP}) and ends the proof of assertion (g). Hence, the proof of the theorem is complete.\end{proof}
\begin{corollary}\label{MainConsequence4Theorem} For any $t\in\{0,1,...,T\}$, the following equalities hold.
\begin{align}
&\underset{\cF_t}{\esssup}(I_{\{\tau\ge t\}})=I_{\{\widetilde{G}_t>0\}},\quad\rm{and}\quad \underset{\cF_t}{\essinf}(I_{\tau\ge t})=I_{\{\widetilde{G}_t=1\}}\label{equality1}\\
& \underset{\cF_t}{ \esssup}(I_{\tau> t})=I_{\{G_t>0\}},\quad\rm{and}\quad\underset{\cF_t}{\essinf}(I_{\tau> t})=I_{\{ G_t=1\}}\label{equality2}\\
& \underset{\cF_{t-1}}{\esssup}(I_{\{\widetilde{G}_t>0\}})=I_{\{G_{t-1}>0\}},\ \rm{and}\quad\underset{\cF_{t-1}}{\essinf}(I_{\{\widetilde{G}_t=1\}})= I_{\{G_{t-1}=1\}} \label{equality3}\\
 & \underset{\cF_{t-1}}{ \esssup}(I_{\{\tau\ge t\}})=I_{\{G_{t-1}>0\}},\quad\rm{and}\quad \underset{\cF_{t-1}}{\essinf}(I_{\{\tau\ge t\}})=I_{\{G_{t-1}=1\}}.\label{equality4}
\end{align}
\end{corollary}
\begin{proof} 1) By considering $X=1_{\{\tau\ge t\}}$ (respectively $X=1_{\{\tau> t\}}$), ${\cal{H}}_2={\cal{G}}_t$, ${\cal{H}}_1={\cal{F}}_t$ and $Y=E[X|{\cal{F}}_t]=\widetilde{G}_t$ (respectively $Y=G_t$), we conclude that both (\ref{equality1})  and (\ref{equality2}) follow from Theorem \ref{ChangeofProbaFiltration}-(a).\\
2) By considering $X=\widetilde{G}_t$, ${\cal{H}}_2={\cal{F}}_t$, ${\cal{H}}_1={\cal{F}}_{t-1}$ and $Y=E[X|{\cal{F}}_{t-1}]={G}_{t-1}$, we deduce that (\ref{equality3}) follows immediately from Theorem \ref{ChangeofProbaFiltration}-(a).\\
3) Thanks to Lemma \ref{TowerProperty4Essup} and both (\ref{equality1}) and (\ref{equality3}),  we derive 
\begin{eqnarray*}
 \underset{\cF_{t-1}}{ \esssup}(I_{\{\tau\ge t\}})= \underset{\cF_{t-1}}{ \esssup}\left( \underset{\cF_{t}}{ \esssup}(I_{\{\tau\ge t\}})\right)= \underset{\cF_{t-1}}{ \esssup}(I_{\{\widetilde{G}_t>0\}})=I_{\{G_{t-1}>0\}}.
\end{eqnarray*}
 This proves the first equality in (\ref{equality4}), while the second equality follows from similar reasoning. This ends the proof of the corollary.\end{proof}
 The following lemma, which is borrowed from Choulli/Deng\cite{ChDen}, plays a central role in the rest of the paper.
\begin{lemma}\label{Qtilde-Definition} The following assertions hold.\\
{\rm{(a)}} For any $t\in\{1,...,T\}$, we have 
\begin{equation}\label{Gtilde4G}
\left(P(\widetilde{G}_t>0|{\cal{F}}_{t-1})>0\right)=\left(G_{t-1}>0\right),\quad P\mbox{-a.s.}.
\end{equation}
{\rm{(b)}} The process $Z^{\mathbb{F}}=(Z^{\mathbb{F}}_t)_{t=0,...,T}$, defined by $Z^{\mathbb{F}}_0=1$ and
\begin{equation}\label{Z(F)}
Z^{\mathbb{F}}_t:=\prod_{s=1}^t\left({{I_{\{\widetilde{G}_s>0\}}}\over{P(\widetilde{G}_s>0|{\cal{F}}_{s-1})}}+I_{\{G_{s-1}=0\}}\right),\quad t=1,...,T,
\end{equation}
is an $\mathbb{F}$-martingale, and hence $\widetilde{Q}:=Z^{\mathbb{F}}_T\cdot{P}$ is a well defined probability measure.\end{lemma}

\begin{corollary} \label{propo-essupQ-P1} Let $\Gamma_t$ be a family of $\cF_t$-measurable random variables, $\widetilde{Q}$ be the probability measure  defined in Lemma \ref{Qtilde-Definition}, and 
\begin{equation}
\widetilde\gamma_t^{\widetilde{Q}}:=\underset{\cF_{t-1}}{\overset{\widetilde{Q}} {\esssup}}(\Gamma_t)\quad\mbox{and}\quad \widetilde\gamma_t:=\underset{\cF_{t-1}}{\esssup}(\Gamma_t{I}_{\{\widetilde{G}_t>0\}}).
\end{equation}
Then the following assertions hold.\\
{\rm{(a)}} If $I_{\{G_{t-1}>0\}}\widetilde\gamma_t^{\widetilde{Q}}\geq 0$ $P$-a.s., then $\widetilde\gamma_t\geq 0$  $P$-a.s., and 
\begin{equation}\label{EssSupQtilde}
I_{\{G_{t-1}>0\}}\widetilde\gamma_t^{\widetilde{Q}}=\widetilde\gamma_t,\quad P\mbox{ -a.s.}.\end{equation}
{\rm{(b)}} If  $\widetilde\gamma_t\geq 0$  $P$-a.s., then we have
$$\left(\widetilde\gamma_t^{\widetilde{Q}}<0\right)\cap(G_{t-1}>0)\subseteq\left(P(\widetilde{G}_t=0<G_{t-1}|{\cal{F}}_{t-1})>0\right),\quad P\mbox{-a.s.}.$$
\end{corollary}
\begin{proof}To prove assertion (a), we apply Theorem \ref{ChangeofProbaFiltration}-(c) to $(\Gamma, Q, {\cal{H}}_1,{\cal{H}}_2)=(I_{\{G_{t-1}>0\}}\Gamma_t,\widetilde{Q},{\cal{F}}_{t-1}, {\cal{F}}_t)$,  $\widetilde{Q}=Z^{\mathbb{F}}_t\cdot P$ on $\cF_t$,  and we derive
$$\underset{{\cal{F}}_{t-1}}{\overset{\widetilde{Q}} {\esssup}} (I_{\{G_{t-1}>0\}}\Gamma_t)=\underset{{\cal{F}}_{t-1}}{\esssup}(I_{\{G_{t-1}>0\}}\Gamma_t{I}_{\{Z^{\mathbb{F}}_t>0\}})\quad P\mbox{-a.s. on}\quad (Z^{\mathbb{F}}_{t-1}>0).$$
As $(Z^{\mathbb{F}}_{t}>0)\cap(G_{t-1}>0)=(\widetilde{G}_t>0)$, we deduce that (\ref{EssSupQtilde}) holds  on the set $(Z^{\mathbb{F}}_{t-1}>0)$. Due to $(Z^{\mathbb{F}}_{t-1}>0)\cap(G_{t-1}>0)=({G}_{t-1}>0)$, we conclude that  $(Z^{\mathbb{F}}_{t-1}=0)\subset(G_{t-1}=0)$. Thus, the two sides of (\ref{EssSupQtilde}) vanish on $(Z^{\mathbb{F}}_{t-1}=0)$, and the conclusion follows. Note that the first statement of (a) is immediate from (\ref{EssSupQtilde}), and the proof of assertion (a) is complete.\\
To prove assertion (b), we suppose that $\widetilde\gamma_t\geq0$ $P$-a.s., and we apply Theorem \ref{ChangeofProbaFiltration}-(g) to the $(\Gamma, Q, {\cal{H}}_1,{\cal{H}}_2)=(\Gamma_t,\widetilde{Q},{\cal{F}}_{t-1}, {\cal{F}}_t)$ afterwards. As a result, we get
$$
\left(\widetilde\gamma_t^{\widetilde{Q}}<0\right)\subseteq\left(P(Z^{\mathbb{F}}_t=0|{\cal{F}}_{t-1})>0\right).
$$
Therefore, assertion (b) follows immediately from combining this latter inclusion with (\ref{Gtilde4G}) and  $(G_{t-1}>0)\cap(Z^{\mathbb{F}}_t=0)=(\widetilde{G}_t=0<G_{t-1})$, which can be easily proved. This ends the proof of the corollary.
\end{proof}
In the remaining part of this subsection, we elaborate the relationship between $\mathbb{G}$-conditional essential supremum and the  $\mathbb{F}$-conditional essential supremum.
\begin{theorem}\label{Essential4Tau}  If $\Gamma\subseteq L^0(\cG_T)$, then the following assertions hold.\\
{\rm{(a)}} For any $t\leq{T}$, we have $P$-a.s.
\begin{equation}\label{inequalities1}
\underset{\cG_{t-1}}{\esssup}(\Gamma I_{\{\tau\ge t\}})\leq \underset{\cF_{t-1}}{\esssup}(\Gamma I_{\{\tau\ge t\}}),\ \mbox{and}\ I_{\{G_{t-1}<1\}} \underset{\cF_{t-1}}{\esssup}(\Gamma I_{\{\tau\ge t\}})\geq 0.\end{equation}
{\rm{(b)}} Let $t\leq{T}$ and $\Sigma_t:= \left( \underset{\cG_{t-1}}{\esssup}(\Gamma I_{\{\tau\ge t\}})\geq 0\right)$. Then $P$-a.s. we get 
\begin{equation}\label{inequalities2}
\begin{split}
\underset{\cG_{t-1}}{\esssup}(\Gamma I_{\{\tau\ge t\}})&=I_{\{\tau\ge t\}}\underset{\cF_{t-1}}{\esssup}(\Gamma I_{\{\tau\ge t\}}),\ \mbox{on}\quad \Sigma_t,\\
 I_{\{G_{t-1}=1\}}\underset{\cF_{t-1}}{\esssup}(\Gamma I_{\{\tau\ge t\}})&=I_{\{G_{t-1}=1\}}\underset{\cG_{t-1}}{\esssup}(\Gamma I_{\{\tau\ge t\}}),\ \mbox{on}\ \Omega\setminus\Sigma_t,\end{split}
\end{equation}
{\rm{(c)}} If $\Gamma\subseteq L^0_+(\cG_T)$ $P$-a.s. and $t\leq T$, then  $P$-a.s. on $(\tau\geq t)$ we have 
\begin{equation} \label{2.7}
\begin{split}
& \underset{\cG_{t-1}}{\esssup}(\Gamma I_{\{\tau\ge t\}})=\underset{\cF_{t-1}}{\esssup}(\Gamma I_{\{\tau\ge t\}}),\\
& {I}_{\{\tau\ge t\}} \underset{\cG_{t-1}}{\essinf}(\Gamma )={I}_{\{\tau\ge t\}}\underset{\cF_{t-1}}{\esssup}\left( \underset{\cG_{t-1}}{\essinf}(\Gamma I_{\{\tau\geq{t}\}})\right).\end{split}\end{equation}
{\rm{(d)}}  For any $t\leq{T}$, $P$-a.s. we have 
\begin{equation} \label{2.7Bis} 
\begin{split}
I_{\{\tau\ge t\}} \underset{\cG_{t-1}}{\esssup}(\Gamma )&=I_{\{\tau\ge t\}} \underset{\cF_{t-1}}{\esssup}(\Gamma ^+I_{\{\tau\ge t\}})+ I_{\{G_{t-1}=1\}} \underset{\cF_{t-1}}{\esssup}(-\Gamma ^-I_{\{\tau\ge t\}})\\
&\hskip 1cm-I_{\{\tau\ge t\}}I_{\{G_{t-1}<1\}}\underset{\cF_{t-1}}{\esssup}\left(\underset{\cG_{t-1}}{\essinf}(\Gamma ^-I_{\{\tau\geq{t}\}})\right). \end{split}\end{equation}
\end{theorem}
The theorem clearly singles out fully the relationship between essential supremum under $\mathbb{G}$ and that under $\mathbb{F}$. In fact, in (\ref{inequalities1}) the theorem states that $\mathbb{F}$-essential supremum is nonnegative on the set $(G_{t-1}<1)$, while it is always an upper bound for $\mathbb{G}$-essential supremum. Thus, we conclude that ${P}$-a.s. on $(G_{t-1}<1)\cap(\underset{\cG_{t-1}}{\esssup}(YI_{\{\tau\ge t\}})<0)$ we have 
\begin{equation*}
 \underset{\cG_{t-1}}{\esssup}(YI_{\{\tau\ge t\}})<0\leq I_{\{\tau\ge t\}}\underset{\cF_{t-1}}{\esssup}(YI_{\{\tau\ge t\}}).
\end{equation*}  
\begin{proof} {\it of Theorem \ref{Essential4Tau}.} The proof of the theorem is divided into four parts, where we prove the four assertions respectively.\\
{\bf Part 1.} Hereto, we prove assertion (a). To this end, we first consider the case where $\Gamma=\{Y\}$ is a singleton. We use the definition of essential supremum, and deduce that $\underset{\cF_{t-1}}{\esssup}(YI_{\{\tau\ge t\}})\in L^0({\cal{F}}_{t-1})\subset L^0({\cal{G}}_{t-1})$ and 
$$ \underset{\cF_{t-1}}{\esssup}(YI_{\{\tau\ge t\}})\geq YI_{\{\tau\ge t\}},\quad P\mbox{-a.s.}.$$
Thus, the first inequality in (\ref{inequalities1}) follows immediately from the above inequality and the definition of essential supremum again. To prove the second inequality in (\ref{inequalities1}), we combine its first inequality with the fact that $ \underset{\cG_{t-1}}{\esssup}(YI_{\{\tau\ge t\}})=I_{\{\tau\ge t\}} \underset{\cG_{t-1}}{\esssup}(Y)$ and get
$
I_{\{\tau<t\}} \underset{\cF_{t-1}}{\esssup}(YI_{\{\tau\ge t\}})\geq 0,$ $P$-a.s.. This is equivalent to $ I_{\{\tau<t\}}\leq I_{\{ \underset{\cF_{t-1}}{\esssup}(YI_{\{\tau\ge t\}})\geq 0\}},\ P\mbox{-a.s}..$ 
Then by taking conditional expectation on both sides of the latter inequality, we derive 
\begin{equation*}1-G_{t-1}\leq  I_{\{ \underset{\cF_{t-1}}{\esssup}(YI_{\{\tau\ge t\}})\geq 0\}},\quad P\mbox{-a.s}..\end{equation*}
Therefore, the second inequality in (\ref{inequalities1})  is a direct consequence of this inequality. This proves (\ref{inequalities1}). For the general case, it suffices to observe that $ \underset{\cF_{t-1}}{\esssup}(\Gamma I_{\{\tau\ge t\}})=\underset{Y\in \Gamma}{\esssup}\, \underset{\cF_{t-1}}{\esssup}(YI_{\{\tau\ge t\}})$ to obtain the same inequalities. This ends the proof of assertion (a).\\
2) This part gives the proof of assertion (b). Thanks to Lemma \ref{Projection2G-F} when $n=1$, we deduce the existence of $\gamma_{t-1}\in L^0(\cF_{t-1})$ such that  
 \begin{equation}\label{F-representation}
 \underset{\cG_{t-1}}{\esssup}(\Gamma 1_{\{\tau\ge t\}})=1_{\{\tau\ge t\}}\underset{\cG_{t-1}}{\esssup}(\Gamma )=1_{\{\tau\ge t\}}\gamma_{t-1},\quad P\mbox{-a.s.}.\end{equation}
As a result, we get $(\underset{\cG_{t-1}}{\esssup}(\Gamma 1_{\{\tau\ge t\}})\geq 0)=(\tau\geq t)\cap (\gamma_{t-1}\geq 0)\cup ( \tau<t)$, and  hence the first equality in (\ref{inequalities2}) is equivalent to prove the equality on $(\tau\geq t)\cap (\gamma_{t-1}\geq 0)$ $P$-a.s. instead. Indeed, the equality is trivial on the set $ ( \tau<t)$.  To this end, we combine (\ref{F-representation}), the tower property of Lemma \ref{TowerProperty4Essup}, and Corollary \ref{MainConsequence4Theorem} and conclude that $P$-a.s. on $(\gamma_{t-1}\geq 0)$ we have 
$$ \esssup_{\cF_{t-1}}(\Gamma 1_{\{\tau\ge t\}})=\esssup_{{\cal{F}}_{t-1}}\left(\underset{{\cal{G}}_{t-1}}{\esssup}(\Gamma I_{\{\tau\ge t\}})\right)=\gamma_{t-1}1_{\{G_{t-1}>0\}}.$$
Then by multiplying both sides with $I_{\{\tau\geq t\}}$, and using $(\tau\geq t)\subset(G_{t-1}>0)$ afterwards, we obtain the first equality in  (\ref{inequalities2}). To prove the second equality in  (\ref{inequalities2}), we notice that in virtue of (\ref{F-representation}), we have 
$$(\underset{\cG_{t-1}}{\esssup}(\Gamma 1_{\{\tau\ge t\}})< 0)=(\tau\geq t)\cap (\gamma_{t-1}< 0),\ \mbox{and}\  (G_{t-1}=1)\subset(\tau\geq t).$$
Hence, the second equality in (\ref{inequalities2}) reduces
$$\underset{\cF_{t-1}}{\esssup}(\Gamma I_{\{\tau\ge t\}})=\underset{\cG_{t-1}}{\esssup}(\Gamma I_{\{\tau\ge t\}}),\ {P}\mbox{-a.s. on}\ ( \gamma_{t-1}< 0)\cap(G_{t-1}=1).$$

Taking the essential supremum knowing $\cF_{t-1}$ on both sides of (\ref{F-representation}), we deduce  by  Corollary \ref{MainConsequence4Theorem} that $P$-a.s. on $(\gamma_{t-1}<0)$ we have 
\begin{equation} \label{Aux3.22} \underset{\cF_{t-1}}{\esssup}(\Gamma 1_{\{\tau\ge t\}})= \gamma_{t-1}\underset{\cF_{t-1}}{\essinf}\left(I_{\{\tau\geq t\}}\right)=\gamma_{t-1}1_{\{G_{t-1}=1\}}.\end{equation}
As $\underset{\cG_{t-1}}{\esssup}(\Gamma I_{\{\tau\ge t\}})=\gamma_{t-1}I_{\{\tau\ge t\}}$ and $I_{\{\tau\ge t\}}I_{\{G_{t-1}=1\}}=I_{\{G_{t-1}=1\}}$, the second equality in  (\ref{inequalities2}) is a direct consequence of (\ref{Aux3.22}). This ends the proof of assertion (b).\\
 3) Herein, we prove assertion (c). To this end we suppose that $\Gamma\subseteq L^0_+({\cal{G}}_T)$, and remark that $ \underset{\cG_{t-1}}{\esssup}(\Gamma I_{\{\tau\ge t\}})\geq \underset{\cG_{t-1}}{\essinf}(\Gamma I_{\{\tau\ge t\}})\geq 0$ $P$-a.s.. Then, in virtue of assertion (b), the first equality in (\ref{2.7}) follows immediately from the first equality in (\ref{inequalities2}), while the second equality of  (\ref{2.7}) is also a direct consequence of the first equality of (\ref{inequalities2})  applied to $\underset{\cG_{t-1}}{\essinf}(\Gamma I_{\{\tau\geq{t}\}})$ instead of $\Gamma$. This completes  the proof of assertion (c).  \\
4) This part deals with assertion (d). Thanks to Lemma \ref{f(esssup)}, we derive  
\begin{equation*}
\begin{split}
I_{\{\tau\ge t\}}\underset{\cG_{t-1}}{\esssup}(\Gamma )&=I_{\{\tau\ge t\}}(\underset{\cG_{t-1}}{\esssup}(\Gamma ))^+ -I_{\{\tau\ge t\}}(\underset{\cG_{t-1}}{\esssup}(\Gamma ))^-\\
&=I_{\{\tau\ge t\}}\underset{\cG_{t-1}}{\esssup}(\Gamma ^+)+I_{\{\tau\ge t\}}\underset{\cG_{t-1}}{\esssup}(-\Gamma ^-)\\
&=\underset{\cG_{t-1}}{\esssup}(\Gamma^+I_{\{\tau\ge t\}})+I_{\{G_{t-1}=1\}}\underset{\cG_{t-1}}{\esssup}(-\Gamma ^-I_{\{\tau\geq t\}})\\
&\quad-I_{\{G_{t-1}<1\}}\underset{\cG_{t-1}}{\essinf}(\Gamma ^-I_{\{\tau\geq t\}}).\end{split}\end{equation*}
Thus, by combining this latter equality with assertions (b) and (c), assertion (d) follows immediately. This ends the proof of theorem.\end{proof}
The second equality in (\ref{2.7}) and its proof convey the following interesting identity.
 \begin{corollary} For any $t\in\{1,..,T\}$ and any $Y\in{L}^0_{+}({\cal{G}}_{t-1})$, we have 
\begin{equation}
YI_{\{\tau\geq{t}\}}=I_{\{\tau\geq{t}\}}\underset{{\cal{F}}_{t-1}}{\esssup}(YI_{\{\tau\geq{t}\}}),\quad P\rm{-a.s.}.
\end{equation}
Or equivalently
\begin{equation}\label{EssSup2ConditionalExpecation}
E\left[YI_{\{\tau\geq{t}\}}\ \big|{\cal{F}}_{t-1}\right]=G_{t-1}\underset{{\cal{F}}_{t-1}}{\esssup}(YI_{\{\tau\geq{t}\}}),\quad P\rm{-a.s.}.\end{equation}
\end{corollary}
This corollary gives more insight about the second equality in Lemma \ref{Projection2G-F}, using essential supremum instead of conditional expectation, for any nonnegative random variable.
\section{Super-hedging prices' set and Immediate-Profit arbitrage}
In this section we will address the pricing method adopted in Carassus-Lepinette \cite{CL} for vulnerable claims. We call vulnerable claims, any claim $H$ that involves the occurrence random time $\tau$ somehow, and hence it is a ${\cal{G}}_T$-measurable random variable and is characterized by a pair of $\mathbb{F}$-adapted processes $(C, R)$. Here $C$ is the payoff process of the claim and $R$ is the recovery process, while the binding relationship between $(C,R)$ and $H$ is dictated by the {\it recovery policy}. The rest of this section is divided into two subsections. The first subsection discusses the pricing set of one-step super-hedging prices for vulnerable claims, while the second subsection elaborates the IP arbitrage results.
\subsection{Super-hedging prices' sets for vulnerable claims}
Throughout the rest of the paper, we consider $(\overline{S},\mathbb{F})$ and $(\widetilde{S},\mathbb{F})$ given by 
\begin{equation}\label{Sbar}
\begin{split}
&\overline{S}:=S_0+\sum_{s=1}^{\cdot}{I}_{\{\widetilde{G}_s>0\}}\Delta{S}_s,\quad 
\widetilde{S}:=S_0+\sum_{s=1}^{\cdot}{I}_{\{G_{s-1}>0\}}\Delta{S}_s,\\
& \Delta{S}_s:=S_s-S_{s-1},\quad s=1,...,T,\quad \sum_{\emptyset}=0.\end{split}\end{equation}
Furthermore, it is easy to see that $\overline{S}_t:=S_0+\sum_{s=1}^t {I}_{\{\widetilde{G}_{s}>0\}}\Delta{\widetilde{S}}_s$. The following theorem constitutes our main result of this subsection, and it explains how the set of super-hedging prices for various vulnerable claims {\it expands} under the effect of the randomness borne in $\tau$. 
\begin{theorem}\label{TheoremSH1-SetOfPrices} Let $t\in\{1,.., T\}$, $\xi\in L^0(\cG_t)$, $(g_s)_{s=0,...,T}$ and $(K_s)_{s=0,....,T}$ be two $\mathbb{F}$-adapted processes, and consider the triplet $(\widehat{g}, \kappa^{(0)}, \kappa^{(g)})$ given by
\begin{equation}\label{gHatKappa}
\begin{split}
\kappa^{(g)}:=\kappa(g,K)&:=gI_{\{\widetilde{G}=G>0\}}+KI_{\{\widetilde{G}>G=0\}}+\max(g,K)I_{\{\widetilde{G}>G>0\}},\\
 \widehat{g}&:=\kappa(g,0),\quad \kappa^{(0)}:=\kappa(0,K),\quad \overline{g}:=gI_{\{\widetilde{G}>0\}}=\kappa(g,g).
\end{split}
\end{equation}
Then the following assertions hold.\\
{\rm{(a)}} If $\xi=g_t1_{\{\tau>t\}}$, then we have 
 \begin{equation}\label{Set4PricesClaim1}\begin{split}
 & \cP_{t-1,t}^{(S^\tau,\mathbb{G})}(\xi)\\
 &=L^0_+(\cG_{t-1})I_{\{\tau\leq t-1\}}+\bigcup_{\delta\in L^0(\cF_{t-1})} \cP_{t-1,t}^{(\overline{S},{\mathbb{F}})}(\widehat g_t+ \delta{I}_{\{\widetilde{G}_t=0<G_{t-1}\}})I_{\{\tau\ge t\}}
\\
&=L^0_+(\cG_{t-1})I_{\{\tau\le t-1\}}+\cP_{t-1,t}^{(\widetilde{S},\mathbb{F},\widetilde Q)}(\widehat{g}_t)I_{\{\tau\ge t\}}.\end{split}\end{equation}
{\rm{(b)}} If $\xi=g_tI_{\{\tau\geq{t}\}}$, then we have 
 \begin{equation}\label{Set4PricesClaim2}\begin{split}
 &  \cP_{t-1,t}^{(S^\tau,\mathbb{G})}(\xi)-L^0_+(\cG_{t-1})I_{\{\tau\leq t-1\}}\\
 &= \bigcup_{\delta\in L^0(\cF_{t-1})} \cP_{t-1,t}^{(\overline{S},{\mathbb{F}})}( \overline{g}_t+ \delta{I}_{\{\widetilde{G}_t=0<G_{t-1}\}})I_{\{\tau\ge t\}}=\cP_{t-1,t}^{(\widetilde{S},\mathbb{F},\widetilde Q)}( \overline{g}_t)I_{\{\tau\ge t\}}.\end{split}\end{equation}
{\rm{(c)}} If $\xi=K_{\tau}I_{\{\tau\le t \}}$, then
\begin{equation}\label{Set4PricesClaim3}
\begin{split}
 & \cP_{t-1,t}^{(S^\tau,\mathbb{G})}(\xi_t)-K_{\tau}I_{\{\tau\le t-1\}}-L^0_+(\cG_{t-1})I_{\{\tau\le t-1\}}\\
 &= \bigcup_{\delta_t\in L^0(\cF_{t-1})}\cP_{t-1,t}^{(\overline{S},\mathbb{F})}(\kappa_t^{(0)}+\delta_{t}I_{\{\widetilde{G}_t=0<G_{t-1}\}})I_{\{\tau\ge t\}}\\
  &=\cP_{t-1,t}^{(\widetilde{S},\mathbb{F},\widetilde{Q})}(\kappa_t^{(0)})I_{\{\tau\ge t\}}.\end{split}\end{equation}
{\rm{(d)}} If $\xi =g_tI_{\{\tau>t\}}+K_{\tau} I_{\{\tau\le t\}}$, then
 \begin{equation}\label{Set4PricesClaim4}  \begin{split}
& \cP_{t-1,t}^{(S^\tau,\mathbb{G})}(\xi)- K_{\tau}I_{\{\tau\le t-1\}}-L^0_+(\cG_{t-1})I_{\{\tau\le t-1\}}\\
 &=\bigcup_{\delta_t\in L^0(\cF_{t-1})} \cP_{t-1,t}^{(\overline{S},\mathbb{F})}(\kappa^{(g)}_t +\delta_tI_{\{\widetilde{G}_t=0<G_{t-1}\}})I_{\{ \tau\ge  t\}},\\
 &= \cP_{t-1,t}^{(\widetilde{S},\mathbb{F},\widetilde{Q})}(\kappa^{(g)}_t )I_{\{\tau\ge  t\}}.\end{split}\end{equation}
 \end{theorem}

No matter what is the vulnerable claim, Theorem \ref{TheoremSH1-SetOfPrices} shows that  {\it this expansion} after $\tau$, i.e. on the set $(\tau<t)$, consists of adding arbitrary nonnegative price, and hence this will have no effect when taking the infinimum. However, for the part before or at $\tau$, i.e. on the set $(\tau\geq{t})$, and again no matter what the vulnerable claim considered, the expansion mechanism of the super-hedging prices' set is obtained, by expanding the set of claims intrinsic to the interplay between $\mathbb{F}$ and $\tau$ described by the pair $(\widetilde{G},G)$. Besides this expansion of claims due to the correlation risks generated by $\tau$, we describe precisely the $\mathbb{F}$-risks which the vulnerable claim entails, and how the set of prices are related. This precise relationship, of quantifying $\mathbb{F}$-risks for the vulnerable claims, is established using the two $\mathbb{F}$-models $(\overline{S},\mathbb{F},P)$ and $(\widetilde{S},\mathbb{F},\widetilde{Q})$.

  \begin{proof}{\it of Theorem \ref{TheoremSH1-SetOfPrices}.} The proof of the theorem is divided into three parts, where we prove assertions (a)-(b), (c)  and (d) respectively.\\
{\bf Part 1:} Herein, we prove assertions (a) and (b). On the one hand, $x_{t-1}^{\mathbb{G}}$ belongs to $\cP_{t-1,t}^{(S^\tau,{\mathbb{G}})}(\xi)$ if and only if there exists $\theta_{t-1}^{\mathbb{G}}\in L^0(\RR^d,{\cal{G}}_{t-1})$ such that 
\bea x_{t-1}^{\mathbb{G}}+\theta_{t-1}^{\mathbb{G}}\Delta S_t^\tau \ge g_t{I}_{\{\tau>t\}},\quad P\rm{-a.s.},\label{SH1}\eea
or equivalently 
\begin{equation}\label{SH2}
I_{\{\tau\le t-1\}}x_{t-1}^{\mathbb{G}}\ge 0,\quad\rm{and}\quad 1_{\{\tau\ge t\}}\left( x_{t-1}^{\mathbb{G}}+\theta_{t-1}^{\mathbb{G}}\Delta S_t\right) \ge g_t1_{\{\tau>t\}}\ P\mbox{-a.s.}.\end{equation}
On the other hand, in  virtue of Lemma \ref{Projection2G-F}, there exists $(x_{t-1}^{\mathbb{F}},\theta_{t-1}^{\mathbb{F}})
$ which belongs to $ L^0({\cal{F}}_{t-1})\times L^0(\RR^d,{\cal{F}}_{t-1})$ satisfying 
$$(x_{t-1}^{\mathbb{F}},\theta_{t-1}^{\mathbb{F}})I_{\{\tau\ge t\}}=(x_{t-1}^{\mathbb{G}},\theta_{t-1}^{\mathbb{G}})I_{\{\tau\ge t\}}.$$
Therefore, by inserting the above equality in (\ref{SH2}), we deduce that $x_{t-1}^{\mathbb{G}}$ belongs to $\cP_{t-1,t}^{(S^\tau,{\mathbb{G}})}(\xi)$ iff there exists $(x_{t-1}^{\mathbb{F}},\theta_{t-1}^{\mathbb{F}})\in L^0({\cal{F}}_{t-1})\times L^0(\RR^d,{\cal{F}}_{t-1})$ such that
\begin{equation}\label{SH3}
\begin{split}
&I_{\{\tau\le t-1\}}x_{t-1}^{\mathbb{G}}\ge 0,\quad{P}\mbox{-a.s.}\quad {I}_{\{\tau= t\}}\left( x_{t-1}^{\mathbb{F}}+\theta_{t-1}^{\mathbb{F}}\Delta S_t\right) \ge 0\quad{P}\mbox{-a.s.},\\
&\rm{and}\quad I_{\{\tau> t\}}\left( x_{t-1}^{\mathbb{F}}+\theta_{t-1}^{\mathbb{F}}\Delta S_t\right) \ge g_t{I}_{\{\tau>t\}}\quad{P}\mbox{-a.s.}.\end{split}\end{equation}
Thanks to Lemma \ref{IneqCharacterizationType2}, the last two equalities above are equivalent to 
\begin{equation} (\widetilde{G}_t-G_t)\left( x_{t-1}^{\mathbb{F}}+\theta_{t-1}^{\mathbb{F}}\Delta\overline{S}_t\right) \ge 0,\quad\mbox{and}\ G_t\left( x_{t-1}^{\mathbb{F}}+\theta_{t-1}^{\mathbb{F}}\Delta\overline{S}_t\right) \geq g_t{G}_t.
\end{equation}
Clearly, these can be rewritten into the following equivalent form of 
\begin{equation}\label{Equivalence1}
x_{t-1}^{\mathbb{F}}+\theta_{t-1}^{\mathbb{F}}\Delta\overline{S}_t\ge g_t{I}_{\{\widetilde{G}_t=G_t>0\}}+g_t^+I_{\{\widetilde{G}_t>G_t>0\}}+\delta_{t-1}1_{\{\widetilde{G}_t=0\}},\end{equation}
for some $\delta_{t-1}\in L^0(\cF_{t-1})$. Hence, we conclude that $x_{t-1}^{\mathbb{G}}\in \cP_{t-1,t}^{(S^\tau,{\mathbb{G}})}(\xi)$ if and only if there exists an ${\cal{F}}_{t-1}$-measurable triplet $(x_{t-1}^{\mathbb{F}},\theta_{t-1}^{\mathbb{F}},\delta_{t-1})$ such that
\bea \nonumber &&I_{\{\tau\le t-1\}}x_{t-1}^{\mathbb{G}}\ge 0,\quad I_{\{\tau\geq t\}}x_{t-1}^{\mathbb{G}}=I_{\{\tau\geq t\}}x_{t-1}^{\mathbb{F}}\\ 
\label{ReverseSH1}&& x_{t-1}^{\mathbb{F}}+\theta_{t-1}^{\mathbb{F}}\Delta\overline{S}_t\ge g_tI_{\{\widetilde{G}_t=G_t>0\}}+g_t^+I_{\{\widetilde{G}_t>G_t>0\}}+\delta_{t-1}I_{\{\widetilde{G}_t=0\}}.
\eea
Or equivalently there exists $(x_{t-1}^{\mathbb{F}},\delta_{t-1})\in \cP_{t-1,t}^{(\overline{S},{\mathbb{F}})}(\xi_t^{\mathbb{F}})\times L^0({\cal{F}}_{t-1})$  such that 
\begin{equation} I_{\{\tau\le t-1\}}x_{t-1}^{\mathbb{G}}\ge 0,\ \rm{and}\ I_{\{\tau\geq t\}}x_{t-1}^{\mathbb{G}}=I_{\{\tau\geq t\}}x_{t-1}^{\mathbb{F}},\ \rm{where}\ \xi_{t}^{\mathbb{F}}:=\widehat{g}_t+\delta_{t-1}I_{\{\widetilde{G}_t=0\}}.
\end{equation}
Therefore, the first equality in (\ref{Set4PricesClaim1}) follows immediately. To prove the second equality, we combine (\ref{SH3}), Lemma \ref{IneqCharacterizationType2} again,  $(\tau\geq t)\subset (G_{t-1}>0)$ and the fact that $\widetilde{Q}(\widetilde{G}_t=0<G_{t-1})=0$, and conclude that   $x_{t-1}^{\mathbb{G}}\in \cP_{t-1,t}^{(S^\tau,{\mathbb{G}})}(\xi)$ iff there exists $(x_{t-1}^{\mathbb{F}},\theta_{t-1}^{\mathbb{F}})\in L^0({\cal{F}}_{t-1})\times L^0(\RR^d,{\cal{F}}_{t-1})$ such that
\begin{equation*}\begin{split}
&I_{\{\tau\le t-1\}}x_{t-1}^{\mathbb{G}}\ge 0\quad P\mbox{-a.s.,}\quad I_{\{\tau\geq t\}}x_{t-1}^{\mathbb{G}}=I_{\{\tau\geq t\}}x_{t-1}^{\mathbb{F}},\quad P\mbox{-a.s.},\\
&\mbox{and}\quad x_{t-1}^{\mathbb{F}}+\theta_{t-1}^{\mathbb{F}}\Delta\widetilde{S}_t\geq \widehat{g}_t,\quad \widetilde{Q}\mbox{-a.s..}
\end{split}
\end{equation*}
This proves the second equality and ends the proof of assertion (a). The proof of assertion (b) mimics exactly the proof of assertion (a) and will be omitted.\\
{\bf Part 2:} Here we prove assertion (c). Suppose that $\xi_t= K_{\tau}1_{\{\tau\le t \}}$. Then $x^{\mathbb{G}}_{t-1}\in\cP_{t-1,t}^{(S^\tau,\mathbb{G})}(\xi_t)$ if and only if there exists $\theta^{\mathbb{G}}_{t-1}\in L^0(\RR^d,\cG_{t-1})$ such that $x_{t-1}^{\mathbb{G}}+\theta_{t-1}^{\mathbb{G}}\Delta S^{\tau}_t\ge K_{\tau}1_{\{\tau\le t \}}$. Again thanks to Lemma \ref{Projection2G-F}, we deduce the existence of a pair  $(x_{t-1}^{\mathbb{F}}, \theta_{t-1}^{\mathbb{F}})\in L^0(\cF_{t-1})\times{ L}^0(\RR^d,\cF_{t-1})$ such that
$$x_{t-1}^{\mathbb{G}}I_{\{\tau \ge t\}}=x_{t-1}^{\mathbb{F}}I_{\{\tau \ge t\}},\quad\rm{ and}\quad \theta_{t-1}^{\mathbb{G}}I_{\{\tau \ge t\}}=\theta_{t-1}^{\mathbb{F}}I_{\{\tau \ge t\}}.$$
 Hence, we deduce that  $x^{\mathbb{G}}_{t-1}\in\cP_{t-1,t}^{(S^\tau,\mathbb{G})}(\xi_t)$ is equivalent to 
\begin{equation} \label{Theo-2ndType-aux1} 
\begin{split}
&(x_{t-1}^{\mathbb{G}}+\theta_{t-1}^{\mathbb{G}}\Delta S_t)I_{\{\tau \ge t\}}=\left(x_{t-1}^{\mathbb{F}}+\theta_{t-1}^{\mathbb{F}}\Delta\overline{ S}_t\right)1_{\{\tau \ge t\}}\ge K_t1_{\{\tau= t \}},\\
&\mbox{and}\quad 
x_{t-1}^{\mathbb{G}}1_{\{\tau \le t-1\}}\ge K_{\tau}1_{\{\tau \le t-1\}} .\end{split}
\end{equation}
Thus, on the one hand, the second  inequality in  (\ref{Theo-2ndType-aux1})  is equivalent to 
\begin{equation}\label{charact1}
x_{t-1}^{\mathbb{G}}1_{\{\tau \le t-1\}}-K_{\tau}1_{\{\tau \le t-1\}}\in{L}^0_+({\cal{G}}_{t-1})I_{\{\tau\leq t-1\}}.\end{equation}
  On the other hand, in virtue of Lemma \ref{IneqCharacterizationType2}, the first inequality in (\ref{Theo-2ndType-aux1}) is equivalent to $x_{t-1}^{\mathbb{F}}+\theta_{t-1}^{\mathbb{F}}\Delta\overline{S}_t\geq  \kappa_t^{(0)}+x_{t-1}^{\mathbb{F}}I_{\{\widetilde{G}_t=0\}},$ or equivalently
  $$x_{t-1}^{\mathbb{F}}\in \bigcup_{\delta_{t-1}\in{L}^0({\cal{F}}_{t-1})}\cP_{t-1,t}^{(\overline{S},\mathbb{F},P)}( \kappa_t^{(0)}+\delta_{t-1}^{\mathbb{F}}I_{\{\widetilde{G}_t=0\}}) .$$
Therefore, by combining this last fact with (\ref{charact1}) and (\ref{Theo-2ndType-aux1}), the first equality in  (\ref{Set4PricesClaim3}) follows immediately, while the proof for the second equality mimics the proof of the second equality in assertion (a) (see part 1). This ends the proof of assertion (c).\\ 
{\bf Part 3.} Hereto, we prove assertion (d). Thus, consider $\xi =g_tI_{\{\tau>t\}}+K_{\tau} I_{\{\tau\le t\}}$. Then  $x^{\mathbb{G}}_{t-1}\in\cP_{t-1,t}^{(S^\tau,\mathbb{G})}(\xi)$ iff there exists $\theta^{\mathbb{G}}_{t-1}\in L^0(\RR^d,\cG_{t-1})$ such that 
\begin{equation}\label{equa100}
x_{t-1}^{\mathbb{G}}+\theta_{t-1}^{\mathbb{G}}\Delta S^{\tau}_t\geq{g}_tI_{\{\tau>t\}}+ K_{\tau}1_{\{\tau\le t \}}\quad P\mbox{-a.s.}.\end{equation}
On the one hand, due to Lemma \ref{Projection2G-F}, there exists a pair $(x_{t-1}^{\mathbb{F}}, \theta_{t-1}^{\mathbb{F}})$, which belongs to $L^0(\cF_{t-1})\times{ L}^0(\RR^d,\cF_{t-1})$ and satisfies $$(x_{t-1}^{\mathbb{G}},  \theta_{t-1}^{\mathbb{G}})I_{\{\tau \ge t\}}=(x_{t-1}^{\mathbb{F}},\theta_{t-1}^{\mathbb{F}})I_{\{\tau \ge t\}}.$$
By inserting these in (\ref{equa100}), we conclude that $x^{\mathbb{G}}_{t-1}\in\cP_{t-1,t}^{(S^\tau,\mathbb{G})}(\xi)$ iff $P$-a.s. $x_{t-1}^{\mathbb{G}}I_{\{\tau\leq{t}-1\}}\geq K_{\tau}I_{\{\tau\leq{t}-1\}}$ and $(x_{t-1}^{\mathbb{F}}+\theta_{t-1}^{\mathbb{F}}\Delta\overline{S}_t)I_{\{\tau\geq{t}\}}\geq{g}_tI_{\{\tau>t\}}+ K_{\tau}1_{\{\tau\le t \}}$. Or equivalently $P$-a.s. we have 
\begin{equation}
\begin{split}
&x_{t-1}^{\mathbb{G}}I_{\{\tau\leq{t}-1\}}\geq{K}_{\tau}I_{\{\tau\leq{t}-1\}},\ (x_{t-1}^{\mathbb{F}}+\theta_{t-1}^{\mathbb{F}}\Delta\overline{S}_t)I_{\{\tau={t}\}}\geq{K}_{t}1_{\{\tau= t \}}\\
&\mbox{and}\ (x_{t-1}^{\mathbb{F}}+\theta_{t-1}^{\mathbb{F}}\Delta\overline{S}_t)I_{\{\tau>{t}\}}\geq{g}_{t}1_{\{\tau> t \}}.
\end{split}
\end{equation}
Thanks to Lemma \ref{IneqCharacterizationType2}-(a)-(b), this equivalent to $P$-a.s. 
\begin{equation}
\begin{split}
&x_{t-1}^{\mathbb{G}}I_{\{\tau\leq{t}-1\}}\geq{K}_{\tau}I_{\{\tau\leq{t}-1\}},\ (x_{t-1}^{\mathbb{F}}+\theta_{t-1}^{\mathbb{F}}\Delta\overline{S}_t)I_{\{G_t>0\}}\geq{g}_{t}1_{\{G_t> 0 \}},\\
&\rm{and}\ (x_{t-1}^{\mathbb{F}}+\theta_{t-1}^{\mathbb{F}}\Delta\overline{S}_t)I_{\{\widetilde{G}_t>G_{t}\}}\geq{K}_{t}1_{\{\widetilde{G}_t>G_{t}\}}.
\end{split}
\end{equation}
Furthermore, it is easy to check that the two last inequalities above are equivalent to 
$
(x_{t-1}^{\mathbb{F}}+\theta_{t-1}^{\mathbb{F}}\Delta\overline{S}_t)I_{\{\widetilde{G}_t>0\}}\geq\kappa^{(g)}_t1_{\{\widetilde{G}_t>0\}}$  $P$-a.s.,or equivalently
$$x_{t-1}^{\mathbb{F}}\in\bigcup_{\delta_t\in{L}^0({\cal{F}}_{t-1})} \cP_{t-1,t}^{(\overline{S},\mathbb{F})}(\kappa^{(g)}_t+\delta_tI_{\{\widetilde{G}_t=0\}}).$$ By combining all these facts, the first equality in (\ref{Set4PricesClaim4}) follows, while the proof of the second equality mimics exactly the proof of the second equality of assertion (a). This proves assertion (d), and ends the proof of theorem.\end{proof}
\subsection{The Immediate-Profit arbitrage under random horizon}
This section analyzes the impact of random horizon on the {\it absence-of-immediate-profit} arbitrage (called AIP hereafter), in many aspects.\\

The following theorem is our first main result of this subsection, and it fully charaterizes the AIP for the model $(S^{\tau},\mathbb{G},P)$  using the model $(\widetilde{S},\mathbb{F},\widetilde{Q}) $.
\begin{theorem} \label{theoAIP-SarreteG} Let $(\overline{S},\mathbb{F},P)$ and $(\widetilde{S},\mathbb{F},\widetilde{Q})$ be the models given by (\ref{Sbar}) and (\ref{Z(F)})-(\ref{Sbar}) respectively, and consider the following assertions.\\
{\rm{(a)}} $(S^{\tau},\mathbb{G},P)$ satisfies the AIP condition\\
{\rm{(b)}}  $(\widetilde{S},\mathbb{F},\widetilde{Q})$ fulfills the AIP condition\\
{\rm{(c)}} $(\overline{S},\mathbb{F},P)$ satisfies the AIP condition.\\
Then (a) $\Longleftrightarrow$ (b) and (b) $\Longrightarrow$ (c).
\end{theorem}
\begin{proof}{\it of Theorem \ref{theoAIP-SarreteG}.} The proof of the theorem will be given in two parts, where we prove (a) $\Longleftrightarrow$ (b) and (b) $\Longrightarrow$ (c) respectively.\\
{\bf Part 1.}  Hereto, we prove (a)$\Longleftrightarrow$ (b). To this end, we suppose that assertion (a) holds. Thus, in virtue of Proposition \ref{GeneralAIP}-(b), assertion (a) is equivalent to 
\begin{equation*}\label{NIP4G}
{\cal{P}}^{(S^{\tau},\mathbb{G})}_{t-1,t}(0)\subseteq L^0_+({\cal{G}}_{t-1}),\quad t\in\{1,...,T\}.
\end{equation*}
Thus, thanks to Theorem \ref{TheoremSH1-SetOfPrices}-(a), these inclusions imply that for $t\in\{1,...,T\}$,
\begin{equation}\label{NIP4G-1}
\begin{split}
{\cal{P}}^{(\widetilde{S},\mathbb{F},\widetilde{Q})}_{t-1,t}(0)I_{\{\tau\geq t\}}&\subseteq{\cal{P}}^{(\widetilde{S},\mathbb{F},\widetilde{Q})}_{t-1,t}(0)I_{\{\tau\geq t\}}+L^0_+({\cal{G}}_{t-1})I_{\{\tau\leq t-1\}}\\
&={\cal{P}}^{(S^{\tau},\mathbb{G})}_{t-1,t}(0)\subseteq L^0_+({\cal{G}}_{t-1}).\end{split}
\end{equation}
Taking conditional expectation knowing $\cF_{t-1}$, we get ${\cal{P}}^{(\widetilde{S},\mathbb{F},\widetilde{Q})}_{t-1,t}(0)G_{t-1}\ge 0$. Therefore, by combining these with ${\cal{P}}^{(\widetilde{S},\mathbb{F},\widetilde{Q})}_{t-1,t}(0)I_{\{G_{t-1}=0\}}\subseteq L^0_+({\cal{F}}_{t-1})$, we deduce that 
${\cal{P}}^{(\widetilde{S},\mathbb{F},\widetilde{Q})}_{t-1,t}(0)\subseteq L^0_+({\cal{F}}_{t-1})\subseteq L^0_+({\cal{F}}_{t-1},\widetilde{Q})$, and assertion (b) follows. This proves  (a)$\Longrightarrow$ (b). To prove the reverse, we assume that assertion (b) holds. In virtue of Proposition \ref{GeneralAIP}-(b), this assumption is equivalent to ${\cal{P}}^{(\widetilde{S},\mathbb{F},\widetilde{Q})}_{t-1,t}(0)\subseteq L^0_+({\cal{F}}_{t-1},\widetilde{Q})$. Then, thanks to Theorem \ref{TheoremSH1-SetOfPrices}, we derive
\begin{equation*}
\begin{split}
{\cal{P}}^{(S^{\tau},\mathbb{G})}_{t-1,t}(0)&=L^0_+({\cal{G}}_{t-1})I_{\{\tau\leq t-1\}}+{\cal{P}}^{(\widetilde{S},\mathbb{F},\widetilde{Q})}_{t-1,t}(0)I_{\{\tau\geq t\}}\\
&\subseteq L^0_+({\cal{G}}_{t-1})I_{\{\tau\leq t-1\}}+L^0_+({\cal{F}}_{t-1})I_{\{\tau\geq t\}}=  L^0_+({\cal{G}}_{t-1}),\end{split}\end{equation*}
where the last equality is a consequence of Lemma \ref{Projection2G-F}. This ends part 1.\\
{\bf Part 2.} This part proves (b) $\Longrightarrow$ (c). Thus, we suppose that assertion (b) holds. Then, in virtue of Proposition \ref{GeneralAIP}-(c), this is equivalent to 
$$
\underset{\cF_{t-1}}{\overset{\widetilde{Q}} {\esssup}}( \theta\Delta\widetilde{S}_t)\ge 0,\quad \widetilde{Q}\mbox{-a.s.},\quad\mbox{for any}\ \ (t,\theta_{t-1})\in\{1,...,T\}\times{L}^0(\RR^d,{\cal{F}}_{t-1}).$$
We deduce that $Z_{t-1}^{\mathbb{F}}\underset{\cF_{t-1}}{\overset{\widetilde{Q}} {\esssup}}( \theta_{t-1}\Delta\widetilde{S}_t)\ge 0$. As $(Z^{\mathbb{F}}_{t-1}>0)\cap(G_{t-1}>0)=({G}_{t-1}>0)$, we get that  $1_{\{G_{t-1}>0\}}\underset{\cF_{t-1}}{\overset{\widetilde{Q}} {\esssup}}( \theta_{t-1}\Delta\widetilde{S}_t)\ge 0$ and  Corollary \ref{propo-essupQ-P1}-(a) applies. Therefore, for any $(t, \theta_{t-1})\in\{1,...,T\}\times{L}^0(\RR^d,{\cal{F}}_{t-1})$, we obtain
$$
\underset{\cF_{t-1}}{\esssup}( \theta_{t-1}\Delta\overline{S}_t)=\underset{\cF_{t-1}}{\esssup}( \theta_{t-1}\Delta\widetilde{S}_tI_{\{\widetilde{G}_t>0\}})\ge 0.$$
Thus, assertion (b) follows from combining this inequality with Proposition \ref{GeneralAIP}-(c). This ends the second part, and completes the proof of the theorem.
\end{proof}
Theorem \ref{theoAIP-SarreteG} gives a complete and full characterization of AIP for the stopped model $(S^{\tau},\mathbb{G})$ using the model $(\widetilde{S},\mathbb{F},\widetilde{Q})$, and shows  that this is sufficient for the AIP fulfillment of the model $(\overline{S},\mathbb{F},P)$. However, in general and in contrast to the classical non-arbitrage (called NA afterwards), the AIP of $(\overline{S},\mathbb{F},P)$ does not give enough information about the AIP for the stopped model. This shows how the impact of $\tau$ can deepen the difference between  AIP and NA, which is pointed out in \cite[Section 2.4]{CL}.
\begin{corollary}\label{Corollary4AIP-Theorem}
If the pair $(S,\tau)$ satisfies 
\begin{equation}\label{Condition4(S,tau)}
P\Bigl(\bigcup_{t=1}^T\Bigl((\Delta{S}_t\not=0)\cap(\widetilde{G}_t=0<G_{t-1})\Bigr)\Bigr)=0,\end{equation}
then the three assertions of Theorem \ref{theoAIP-SarreteG} are equivalent. More precisely, \\
\begin{equation*}\label{3Equivalneces}
(\widetilde{S},\mathbb{F},\widetilde{Q})\ \mbox{ fulfills AIP iff}\ (\overline{S},\mathbb{F},P)\ \mbox{fulfills AIP  iff}\ ({S}^{\tau},\mathbb{G},P)\ \mbox{ satisfies AIP}.\end{equation*}
\end{corollary}
\begin{proof}
Thanks to Theorem  \ref{theoAIP-SarreteG}, the proof of this corollary boils down to prove that, under (\ref{Condition4(S,tau)}), the AIP of $(\overline{S},\mathbb{F},P)$ implies the AIP of $(\widetilde{S},\mathbb{F},\widetilde{Q})$. To this end,  note that $\Delta \widetilde S_t=I_{\{G_{t-1}>0\}}\Delta S_t=I_{\{G_{t-1}>0\}}\Delta S_t I_{\{ \Delta S_t\ne 0\}}$. By assumption, $I_{\{G_{t-1}>0\}}I_{\{ \Delta S_t\ne 0\}}=I_{\{ \Delta S_t\ne 0\}}1_{\{G_{t-1}>0\}}I_{\{\widetilde G_t>0\}}$. Due to $(\widetilde G_t>0)\subseteq (G_{t-1}>0)$, we get $\Delta\widetilde S_t=I_{\{\widetilde G_t>0\}}\Delta S_t=\Delta \overline{S}_t$, i.e. $\widetilde S=\overline{S}$. Thus, by combining this with the fact that $\widetilde{Q}$ is equivalent to $P$ on $(G_{t-1}>0)$ which is due to $(Z^{{\mathbb{F}}_{t-1}}>0)\cap(G_{t-1}>0)=(G_{t-1}>0)$, we easily derive 
$$
\underset{\cF_{t-1}}{\esssup}( \theta_{t-1}\Delta\overline{S}_t)=\underset{\cF_{t-1}}{\esssup}( \theta_{t-1}\Delta\widetilde{S}_t)=\underset{\cF_{t-1}}{\overset{\widetilde{Q}} {\esssup}}( \theta_{t-1}\Delta\widetilde{S}_t),\quad P\mbox{-a.s.}.
$$
for any $\theta\in L^0(\RR^d,{\cal{F}}_{t-1})$. Therefore, the conclusion follows immediately from these equalities.
\end{proof}
\begin{remark}{\rm{(a)}} If the random time $\tau$ satisfies
\begin{equation}\label{Condition4Tau}
P\Biggl(\bigcup_{t=1}^T\bigl\{\widetilde{G}_t=0<G_{t-1}\bigr\}\Biggr)>0,
\end{equation}
then there exist models for $S$ such that  $(\overline{S},\mathbb{F})$ satisfies AIP, while $(S^{\tau},\mathbb{G})$ violates it. In fact, consider 
$$\Delta{S}_t:=I_{\{\widetilde{G}_t=0<G_{t-1}\}}-P(\widetilde{G}_t=0<G_{t-1}\big|{\cal{F}}_{t-1}),\quad t=1,..., T,$$
and derive, by Theorem \ref{ChangeofProbaFiltration}, the following 
\begin{equation*}\begin{split}
&\underset{\cF_{t-1}}{\esssup}( \theta\Delta\overline{S}_t)=\theta^-P(\widetilde{G}_t=0<G_{t-1}\big|{\cal{F}}_{t-1})\geq 0,\ \mbox{for any}\ \theta\in{L}^0({\cal{F}}_{t-1}),\\
&\underset{\cF_{t-1}}{\overset{\widetilde{Q}} {\esssup}}( \theta\Delta\widetilde{S}_t)=-\theta{P}(\widetilde{G}_t=0<G_{t-1}\big|{\cal{F}}_{t-1}),\ \widetilde{Q}\mbox{-a.s.},\ \mbox{for any}\ \theta\in{L}^0({\cal{F}}_{t-1}),\\\
&\underset{\cG_{t-1}}{\overset{P} {\esssup}}( \theta\Delta{S^{\tau}}_t)=-\theta{P}(\widetilde{G}_t=0<G_{t-1}\big|{\cal{F}}_{t-1})I_{\{\tau\geq{t}\}},\ P\mbox{-a.s.},\ \mbox{for}\ \theta\in{L}^0({\cal{G}}_{t-1}). 
\end{split} 
\end{equation*}
Thus, it is clear that, the first inequality above proves that $(\overline{S},\mathbb{F})$ fulfills AIP, while the second and third equalities prove that both $(\widetilde{S},\mathbb{F},\widetilde{Q})$ and $(S^{\tau},\mathbb{G},P)$ violate AIP respectively. \\
{\rm{(b)}} Suppose that $\tau$ satisfies (\ref{Condition4Tau}), and consider the model for $S$ given by $\Delta{S}_t:=I_{\{\widetilde{G}_t=0\}}$, $t\in\{1,...,T\}$ and $S_0=1$. Then $(S,\mathbb{F})$ violates the AIP and hence violates NA also, due to 
$$
\underset{\cF_{t-1}}{\esssup}( \theta_{t-1}\Delta{S}_t)=\theta_{t-1}^+{I}_{\{P(\widetilde{G}_t=0|{\cal{F}}_{t-1})>0\}}-\theta_{t-1}^-{I}_{\{G_{t-1}=0\}}.$$
Furthermore, we have $S^{\tau}=S_0$ and hence $(S^{\tau},\mathbb{G})$ fulfills AIP and NA. \\
{\rm{(c)}} Suppose that $\tau$ satisfies (\ref{Condition4Tau}), and consider $S$ given by $\Delta{S}_t:=I_{\{\widetilde{G}_t=0<G_{t-1}\}}$, $t\in\{1,...,T\}$ and $S_0=1$. Therefore, one can check that $(S,\mathbb{F})$ fulfills AIP and violates NA. In fact, there exists no positive $\mathbb{F}$-martingale $Z$ such that 
$$
E\left[\Delta{S}_t{{Z_t}\over{Z_{t-1}}}\big|{\cal{F}}_{t-1}\right]=0,\ \mbox{and}\ \underset{\cF_{t-1}}{\esssup}( \theta\Delta{S}_t)=\theta^+{I}_{\{P(\widetilde{G}_t=0<G_{t-1}|{\cal{F}}_{t-1})>0\}}\geq 0,$$
for any $\theta\in{L}^0_+({\cal{F}}_{t-1})$. However, we have $S^{\tau}=S_0$ and, as a consequence, $(S^{\tau},\mathbb{G})$ fulfils NA  and hence it satisfies AIP also.
\end{remark}
Our second main theorem, of this subsection, describes the models of $\tau$ for which the AIP condition is unaffected after stopping. 
\begin{theorem}\label{UniversalNIP} The following assertions are equivalent.\\
{\rm{(a)}}  For any $(X,\mathbb{F},P)$ satisfying AIP, the model $(X^{\tau}, \mathbb{G}, P)$ fulfills AIP.\\
{\rm{(b)}} For any $t\in\{1,...,T\}$, we have $\{G_{t-1}=0\}=\{\widetilde{G}_t=0\}$ $P$-a.s..\\
{\rm{(c)}} The probability $\widetilde{Q}$ defined in (\ref{Z(F)}) coincides with $P$, i.e. $Z^{\mathbb{F}}\equiv 1$.
\end{theorem}
The proof of this theorem is based on the following simple but useful lemma.
\begin{lemma} \label{propo-AIP-predictable} Consider any model $(X, \mathbb{H}:=(\cH_t)_{t=0,...,T})$. Then the following assertions hold.\\
{\rm{(a)}} Suppose that $X$ is $\mathbb{H}$-predictable.  Then $(X, \mathbb{H})$ satisfies AIP if and only if $X$ is a constant process, i.e., $X_t=X_0$, $P$-a.s. for any $t=1,...,T$.\\
{\rm{(b)}} $(X, \mathbb{H})$ satisfies AIP if and only if $(\psi\is X, \mathbb{H})$ fulfills AIP, for any $\mathbb{H}$-predictable and bounded process $\psi$, where $\psi\is X_t:= \sum_{s=1}^t\psi_s\Delta{X}_s$, $t\in\{1,...,T\}$.
\end{lemma}
\begin{proof}
1)  If $X$ is a constant process, then it is clear that ${\cal{P}}_t^{(X,\mathbb{H})}(0)=L^0_+({\cal{H}}_t)$. This implies that ${\cal{P}}_{t}^{(X,\mathbb{H})}(0)\cap{L}^0_-({\cal{H}}_t)=\{0\}$, or equivalently AIP holds for $(X,\mathbb{H})$. To prove the reverse sense, we suppose ${\cal{P}}_{t}^{(X,\mathbb{H})}(0)\cap{L}^0_-({\cal{H}}_t)=\{0\}$, and consider any $t\in\{1,..,T\}$ and  $\theta\in{L}^0({\cal{H}}_{t-1})$. Therefore, by combining all these properties, we derive 
$$
\underset{{\cal{H}}_{t-1}}\esssup(-\theta_{t-1}\Delta{X}_t)=-\theta_{t-1}\Delta{X}_t\in {\cal{P}}_{t-1}^{(X,\mathbb{H})}(0)\subset{L}^0_+({\cal{H}}_{t-1}).$$ 
Thus, as $\theta_{t-1}$ is arbitrary, we conclude that $\theta_{t-1}\Delta{X}_t=0$ $P$-a.s., for any $\theta_{t-1}\in{L}^0({\cal{H}}_{t-1})$. This implies that $\Delta{X}_t=0$ $P$-.a.s., and the proof of assertion (a)  is complete.\\
2) If $(\psi\is X, \mathbb{H})$ fulfills AIP, for any $\mathbb{H}$-predictable and bounded process $\psi$, then by taking $\psi=1$ we deduce that $(X-X_0,\mathbb{H})$ satisfies AIP or equivalently $(X,\mathbb{H})$ satisfies AIP.  The reverse sense is a direct consequence from combining Proposition \ref{GeneralAIP}-(b) and the fact that, for any $t\in\{1,...,T\}$ and any $\mathbb{H}$-predictable process $\psi$, we have
$$\underset{\theta_{t-1}\in{L}^0({\cal{H}}_{t-1})}{\essinf}\underset{{\cal{H}}_{t-1}}{\esssup}(\theta_{t-1}\Delta{X}_t)\leq \underset{\varphi_{t-1}\in{L}^0({\cal{H}}_{t-1})}{\essinf}\underset{{\cal{H}}_{t-1}}{\esssup}(\varphi_{t-1}\psi_t\Delta{X}_t),\quad P\mbox{-a.s..}$$
This ends the proof of the lemma.
\end{proof}
\begin{proof}{\it of Theorem \ref{UniversalNIP}.} The equivalence between assertion (b) and (c) can be found in Choulli/Deng \cite{ChDen}, and for the sake of completeness we reproduce it here. Thus, remark that  $Z^{\mathbb{F}}\equiv 1$ --i.e. for any $t\in\{1,...,T\}$ we have ${{Z^{\mathbb{F}}_t/ Z^{\mathbb{F}}_{t-1}}}=1$ $P$-a.s.-- is equivalent to 
$${{I_{\{\widetilde{G}_t>0\}}}\over{P(\widetilde{G}_t>0|{\cal{F}}_{t-1})}}=I_{\{G_{t-1}>0\}},\quad P-a.s.,$$
or equivalently, 
$$I_{\{\widetilde{G}_t>0\}}=P(\widetilde{G}_t>0|{\cal{F}}_{t-1})I_{\{G_{t-1}>0\}}=P(\widetilde{G}_t>0|{\cal{F}}_{t-1}),\quad P\mbox{-a.s.}.$$
Therefore, this is equivalent to $\{P(\widetilde{G}_t>0|{\cal{F}}_{t-1})=1\}=\{\widetilde{G}_t>0\}$, while the latter equality holds if and only if $\{G_{t-1}>0\}=\{G_{t}>0\}$. Hence, (b) $\Longleftrightarrow$ (c) is proved, and the rest of this proof proves  (b) $\Longleftrightarrow$ (a).\\
 Suppose that assertion (b) holds. Then we get $(\overline{X},\mathbb{H},P)=(\widetilde{X},\mathbb{H},P)$ and it satisfies AIP for any model $(X,\mathbb{H})$ satisfying AIP, see Lemma \ref{propo-AIP-predictable} (b). Thus, in virtue of Theorem \ref{theoAIP-SarreteG} , $(X^{\tau},\mathbb{G})$ satisfies AIP. This proves the implication (b) $\Longrightarrow$ (a). To prove the reverse, assume assertion (a) holds, and consider
$$X_t:=\sum_{s=1}^t\left(I_{\{\widetilde{G}_s=0\}}-P(\widetilde{G}_s=0\big|{\cal{F}}_{s-1})\right),\quad t=1,...,T.$$
Then it is clear that $(X,\mathbb{F})$ satisfies NA and a fortiori AIP due to Carassus-Lepinette \cite{CL}. Due to assertion (a), the AIP for $(X^{\tau},\mathbb{G})$ follows, and due to $(\tau\geq s)\cap(\widetilde{G}_s=0)=\emptyset$, we get $X^{\tau}=-\sum_{s=1}^t{I}_{\{s\leq\tau\}}P(\widetilde{G}_s=0\big|{\cal{F}}_{s-1})$ and hence it is $\mathbb{G}$-predictable. Thus, in virtue of Lemma \ref{propo-AIP-predictable}-(a), we conclude that $X^{\tau}$ is a null process, or equivalently for any $t\in\{1,...,T\}$ 
$$
{I}_{\{t\leq\tau\}}P(\widetilde{G}_t=0\big|{\cal{F}}_{t-1})=0,\quad P\mbox{-a.s.}.$$
By taking conditional expectation with respect to ${\cal{F}}_{t-1}$ on both sides of this equality, we get  $G_{t-1}P(\widetilde{G}_t=0\big|{\cal{F}}_{t-1})=0,$ $P$-a.s., for any $t\in\{1,...,T\}$. 
This yields $(G_{t-1}>0)\cap(\widetilde{G}_t=0)=\emptyset$, $P$-a.s. for any $t\in\{1,...,T\}$ and assertion (b) follows immediately. This ends the proof of the theorem.\end{proof}
\section{Pricing formulas for vulnerable claims}
In this section, we elaborate super-hedging-pricing formulas for several vulnerable claims. Our pricing formulas relies on \cite[Lemma 3.1]{CL}, which we recall below, and which elaborates a backward equation for the super-hedging prices for an arbitrary model $(X,\mathbb{H},Q)$.
\begin{lemma}\label{OnePerdiod2MultiPeriod} Let $(X,\mathbb{H},Q)$ be an arbitrary model satisfying AIP and  defined on the probability space $(\Omega,{\cal{G}},P)$, $T\in(0,\infty)$  is a fixed investment horizon, and $\xi\in{L}^0({\cal{H}}_T)$ is a claim. Then the super-hedging price process for $\xi$, denoted by $\widehat{\cP}^{\mathbb{H}}$, is given by the following backward formula.
\begin{equation}\label{PriceAlgorithm}
\begin{split}\widehat{\cP}_t^{\mathbb{H}}&=\underset{\theta\in{L}^0({\cal{H}}_{t})}{\essinf}\underset{{\cal{H}}_{t}}{\overset{Q}{\esssup}}\left(\widehat{\cP}_{t+1}^{\mathbb{H}}-\theta\Delta{X}_{t+1}\right)=\widehat{\cP}_{t,t+1}^{\mathbb{H}}\left(\widehat{\cP}_{t+1}^{\mathbb{H}}\right),\ t\leq{T-1},\\
 \widehat{\cP}_T^{\mathbb{H}}&=\xi.\end{split}
\end{equation}
\end{lemma}
 The vulnerable claims that we address herein can be classified into two main classes. The first class consists of vulnerable claims that do not have recovery, or equivalently there is no payment at the random time. The second class of vulnerable claims is those claims which involve payment at the random time in a way or another.
In virtue of Lemma \ref{OnePerdiod2MultiPeriod}, an important step in  describing the  $\mathbb{G}$-price process for vulnerable claims lies in addressing the impact of $\tau$ on the one-step pricing operator $\widehat{\cP}_{t,t+1}^{(X,\mathbb{H},Q)}(\cdot)$, defined in (\ref{infinimumPrice}), for any model $(X,\mathbb{H},Q)$. \\

The remaining part of this section is divided into three subsections. In the first subsection, we outline the main results on the one-step pricing operators. The second subsection elaborates the general pricing formulas, while the last subsection proves the main theorems of the second subsection. 
 \subsection{The one-step pricing formulas}
In this subsection, we address the one-step-pricing operator under the random horizon $\tau$ in many aspects. Precisely, following the same spirit as in arbitrage theory, we aim to understand how the one-step-pricing under $\mathbb{G}$ can be described using $\mathbb{F}$-observable processes and pricing-operators. 
\begin{theorem}\label{TheoremSH1-MinimalPrices} Let $t\in\{1,.., T\}$, $\xi\in L^0(\cG_t)$, $(g_s)_{s=0,...,T}$ and $(K_s)_{s=0,....,T}$ be two $\mathbb{F}$-adapted processes, and consider 
$(\widehat{g},\kappa^{(0)},\kappa^{(g)}, \overline{g})$ given by (\ref{gHatKappa}).Then the following assertions hold.\\
{\rm{(a)}} If $\xi=g_t I_{\{\tau>t\}}$, then we have 
 \begin{eqnarray} \label{Price-Claim1}\widehat \cP_{t-1,t}^{(S^\tau,\mathbb{G})}(\xi)=\underset{\delta\in{L}^0({\cal{F}}_{t-1})}{\essinf}\widehat \cP_{t-1,t}^{(\overline{S},\mathbb{F})}(\widehat{g}_t+\delta{I}_{\{\widetilde{G}_t=0\}})I_{\{\tau\ge t\}}
 =\widehat \cP_{t-1,t}^{(\widetilde{S},\mathbb{F},\widetilde Q)}(\widehat g_t )I_{\{\tau\ge t\}}.\quad\quad\label{Equality12}\end{eqnarray}
{\rm{(b)}} If $\xi=g_tI_{\{\tau\geq{t}\}}$, then we have 
 \begin{eqnarray} \label{Price-Claim2}\widehat \cP_{t-1,t}^{(S^\tau,\mathbb{G})}(\xi)=\underset{\delta\in{L}^0({\cal{F}}_{t-1})}{\essinf}\widehat \cP_{t-1,t}^{(\overline{S},\mathbb{F})}( \overline{g}_t+\delta{I}_{\{\widetilde{G}_t=0\}})I_{\{\tau\ge t\}}=\widehat \cP_{t-1,t}^{(\widetilde{S},\mathbb{F},\widetilde Q)}( \overline{g}_t )I_{\{\tau\ge t\}}.\quad\quad\label{Equality22}\end{eqnarray}
{\rm{(c)}} If $\xi=K_{\tau}I_{\{\tau\le t \}}$, then
\begin{equation} \label{Min des prix-Type2} 
\begin{split}
  \widehat \cP_{t-1,t}^{(S^\tau,\mathbb{G})}(\xi_t)&=K_{\tau} I_{\{\tau\le t-1 \}}+\underset{\delta_t\in{L}^0({\cal{F}}_{t-1})}{\essinf} \widehat \cP_{t-1,t}^{(\overline{S},\mathbb{F})}(\kappa_t^{(0)}+\delta_t{I}_{\{\widetilde{G}_t=0\}})I_{\{\tau\ge t\}}\\
  &=K_{\tau} I_{\{\tau\le t-1 \}}+ \widehat \cP_{t-1,t}^{(\widetilde{S},\mathbb{F},\widetilde{Q})}(\kappa_t^{(0)} )I_{\{\tau\ge t\}}.\end{split}
  \end{equation}
{\rm{(d)}} If $\xi =g_tI_{\{\tau>t\}}+K_{\tau} I_{\{\tau\le t\}}$, then
 \begin{equation}\label{Price-Claim3} 
 \begin{split}
 \widehat \cP_{t-1,t}^{(S^\tau,\mathbb{G})}(\xi)&=K_{\tau} I_{\{\tau\le t-1 \}}+\underset{\delta\in{L}^0({\cal{F}}_{t-1})}{\essinf}\widehat \cP_{t-1,t}^{(\overline{S},\mathbb{F})}(\kappa_t^{(g)} +\delta{I}_{\{\widetilde{G}_t=0\}})I_{\{\tau> t-1\}}\\
 &=K_{\tau} I_{\{\tau\le t-1 \}}+\widehat \cP_{t-1,t}^{(\widetilde{S},\mathbb{F},\widetilde{Q})}(\kappa_t^{(g)})I_{\{\tau\geq{t}\}}.\end{split}\end{equation}
  \end{theorem}
 This theorem quantifies the one-step-pricing operator for $(S^{\tau},\mathbb{G})$ in terms of pricing operators for both models $(\overline{S},\mathbb{F},P)$ and $(\widetilde{S},\mathbb{F},\widetilde{Q})$. However, the latter model is the one that gives us full $\mathbb{F}$-characterization. Assertion (a) addresses the first class of vulnerable claims, where there is no recovery at all (i.e. no payment at the random time $\tau$ when it occurs). In life insurance, this claim belongs to the class when benefit is paid upon survival only. Assertion (c) deals with vulnerable claims that have payments at the random time only, and this also has a meaning in life insurance, which consists of getting benefit when the insured dies and nothing in case of survival. Assertion (d) treats another vulnerable claim with recovery, as it combines both previous cases by paying benefits in both situation when the insured survives and when she dies. The claim addressed in assertion (b) is somehow in between the two classes. On the one hand, mathematically, it belongs to the second class by choosing $K=g_TI_{\Rbrack{T}\Lbrack}$ as there is payment at $\tau$ which coincides with the payment upon survival. On the other hand, its pricing formula in (\ref{Price-Claim2}) tells us that this claim falls into the first class immediately after the one-step-pricing.
\begin{remark} \label{Remark4ClaimsOneStepPricing}  
{\rm{(i)}} Assertion (a) conveys that the one-step super-hedging price of the claim having no recovery (i.e. no payment at $\tau$ at all) has literally the same form. This is due to the fact that $(\tau\geq t)=(\tau>t-1)$.  In other words, by denoting ${\widehat{\cP}}_{\mbox{one-step}}(\cdot)$ the one-step super-hedging pricing operator, then ${\widehat{\cP}}_{\mbox{one-step}}(\mbox{Class1})\subseteq \mbox{Class1}$, and ${\widehat{\cP}}_{\mbox{one-step}}(\mbox{claim-of-assertion-(b)})\in \mbox{Class1}$.\\
{\rm{(ii)}} Assertion (d) proves that the one-step super-hedging price of the claim has exactly the same form as the claim, while the one-step super-hedging pricing for the claim of assertion (c) transforms the claim into the claim of assertion (d). In other words, even though there is no payment upon the survival, after one-step super-hedging pricing the claim will have payment upon survival.
\end{remark}
Theorem \ref{TheoremSH1-MinimalPrices} conveys also, via the formulas, that the one-step pricing operators for $(\widetilde{S},\mathbb{F},\widetilde{Q})$ and  $(\overline{S},\mathbb{F},P)$ might differ in general, while the resulting prices for any claim are comparable. In the following remark, we discuss these points in details.  
\begin{remark}
{\rm{(a)}}  For any $\delta\in L^0(\cF_{t-1})$, the following inequalities hold
\begin{equation}\label{SecondEquality}
\begin{split}
&\widehat \cP_{t-1,t}^{(\overline{S},{\mathbb{F}})}(\xi^{\mathbb{F}} +\delta{I}_{\{\widetilde{G}_t=0\}})I_{\{P(\widetilde{G}_t>0|{\cal{F}}_{t-1})=1\}}= \widehat \cP_{t-1,t}^{(\overline{S},{\mathbb{F}})}(\xi^{\mathbb{F}} )I_{\{P(\widetilde{G}_t>0|{\cal{F}}_{t-1})=1\}},\\
& \widehat \cP_{t-1,t}^{(\overline{S},{\mathbb{F}})}(\xi^{\mathbb{F}} +\delta{I}_{\{\widetilde{G}_t=0\}}){I}_{\{P(\widetilde{G}_t=0|{\cal{F}}_{t-1})>0\}}\geq\delta{I}_{\{P(\widetilde{G}_t=0|{\cal{F}}_{t-1})>0\}}.\end{split}\end{equation}
{\rm{(b)}} For $Y\in\{\widehat{g},\widetilde{g},\overline{g},\kappa\}$, and any $t\in\{1,...,T\}$, we have 
\begin{equation}
\widehat \cP_{t-1,t}^{(\widetilde{S},\mathbb{F},\widetilde Q)}(Y_t )\leq \widehat \cP_{t-1,t}^{(\overline{S},\mathbb{F})}(Y_t),\quad P\mbox{-a.s. on}\quad \{G_{t-1}>0\}.
\end{equation}
As a result, this gives us another proof for Theorem \ref{theoAIP-SarreteG}  using this inequality  and Proposition \ref{GeneralAIP}-(d).
\end{remark}
 In most of applications, $K$ represents the recovery process and it is usually nonnegative. Thus, as a particular case, we consider the case of {\it{vulnerable options}} where both payoff process $g$ and recovery process $K$ are nonnegative. 
\begin{corollary}\label{case4Options} Consider the notation of Theorem \ref{TheoremSH1-MinimalPrices}, and assume that  both processes $g$ and $K$ are nonnegative.  Then the following assertions hold.\\
{\rm{(a)}} If $\xi^{(1)}:=g_t I_{\{\tau>t\}}$ and $\xi^{(2)}:=g_tI_{\{\tau\geq{t}\}}$, then we have 
 \begin{equation} \label{Price-Claim1_Options}
 \begin{split}
 \widehat \cP_{t-1,t}^{(S^\tau,\mathbb{G})}(\xi^{(1)})&=\underset{\delta\in{L}^0({\cal{F}}_{t-1})}{\essinf}\widehat \cP_{t-1,t}^{(\overline{S},\mathbb{F})}({g}_t{I}_{\{G_t>0\}}+\delta{I}_{\{\widetilde{G}_t=0\}})I_{\{\tau\ge t\}}\\
& =\widehat \cP_{t-1,t}^{(\widetilde{S},\mathbb{F},\widetilde Q)}(g_t{I}_{\{G_t>0\}})I_{\{\tau\ge t\}},\\
 \widehat \cP_{t-1,t}^{(S^\tau,\mathbb{G})}(\xi^{(2)})&=\underset{\delta\in{L}^0({\cal{F}}_{t-1})}{\essinf}\widehat \cP_{t-1,t}^{(\overline{S},\mathbb{F})}(\overline{g}_t+\delta{I}_{\{\widetilde{G}_t=0\}})I_{\{\tau\ge t\}}\\
& =\widehat \cP_{t-1,t}^{(\widetilde{S},\mathbb{F},\widetilde Q)}(\overline{g}_t)I_{\{\tau\ge t\}}\end{split} 
 \end{equation}
{\rm{(b)}} If $\xi=K_{\tau}I_{\{\tau\le t \}}$, then
\bea \label{Min des prix-Type2_Options}   \widehat \cP_{t-1,t}^{(S^\tau,\mathbb{G})}(\xi_t)=K_{\tau} I_{\{\tau\le t-1 \}}+ \widehat \cP_{t-1,t}^{(\widetilde{S},\mathbb{F},\widetilde{Q})}(K_t{I}_{\{\widetilde{G}_t>G_t\}} )I_{\{\tau\ge t\}}.\eea
{\rm{(c)}} If $\xi =g_tI_{\{\tau>t\}}+K_{\tau} I_{\{\tau\le t\}}$, then
 \begin{equation}\label{Price-Claim3_Options}
  \widehat \cP_{t-1,t}^{(S^\tau,\mathbb{G})}(\xi)=K_{\tau} I_{\{\tau\le t-1 \}}+\widehat \cP_{t-1,t}^{(\widetilde{S},\mathbb{F},\widetilde{Q})}\bigl(\max(g_t{I}_{\{G_t>0\}},K_t{I}_{\{\widetilde{G}_t>G_t\}})\bigr )I_{\{\tau\geq{t}\}}.
 \end{equation}
\end{corollary}
The proof of the corollary follows from previous results, and will be omitted.
\begin{proof}{\it of Theorem \ref{TheoremSH1-MinimalPrices}.}
The proof of the theorem will be achieved in three parts. The first and second parts prove two roughly general claims, while the third part outlines the proof for the theorem. To this end, we fix $t\in\{1,...,T\}$, and we consider a triplet $(\Xi, \xi^{\mathbb{G}},\xi^{\mathbb{F}})\in{L}^0({\cal{G}}_{t-1})\times{L}^0({\cal{G}}_t)\times{L}^0({\cal{F}}_t)$ such that $\xi^{\mathbb{F}}I_{\{\widetilde{G}_t=0\}}=0$ $P$-a.s..\\
{\bf Part 1.} Herein, we prove that the equality
\begin{equation}\label{Equlity4Sets}
\cP_{t-1,t}^{(S^\tau,{\mathbb{G}})}(\xi^{\mathbb{G}})=\Xi+L^0_+({\cal{G}}_{t-1})I_{\{\tau\leq{t}-1\}}+ \bigcup_{\delta\in{L}^0({\cal{F}}_{t-1})}\cP_{t-1,t}^{(\overline{S},{\mathbb{F}})}(\xi^{\mathbb{F}} +\delta{I}_{\{\widetilde{G}_t=0\}})I_{\{\tau\ge t\}},\end{equation}
always implies
\begin{equation}\label{FirstEquality}
 \widehat \cP_{t-1,t}^{(S^\tau,{\mathbb{G}})}(\xi^{\mathbb{G}})=\Xi+\underset{\delta\in L^0(\cF_{t-1})}\essinf\widehat \cP_{t-1,t}^{(\overline{S},{\mathbb{F}})}(\xi^{\mathbb{F}} +\delta{I}_{\{\widetilde{G}_t=0\}})I_{\{\tau\ge t\}}.
\end{equation}
 To this end, we remark that (\ref{Equlity4Sets}) implies that 
\begin{equation*}
\begin{split}
&\Xi+ \cP_{t-1,t}^{(\overline{S},{\mathbb{F}})}(\xi^{\mathbb{F}} +\delta{I}_{\{\widetilde{G}_t=0\}})I_{\{\tau\ge t\}}\\
&\subset\Xi+L^0_+({\cal{G}}_{t-1})I_{\{\tau\leq{t}-1\}}+ \cP_{t-1,t}^{(\overline{S},{\mathbb{F}})}(\xi^{\mathbb{F}} +\delta{I}_{\{\widetilde{G}_t=0\}})I_{\{\tau\ge t\}}\subset \cP_{t-1,t}^{(S^\tau,{\mathbb{G}})}(\xi^{\mathbb{G}}),\end{split}\end{equation*}
for any $\delta\in{L}^0({\cal{F}}_{t-1})$. Hence, after taking essential infimum, this inclusion implies that  
\begin{equation}\label{Inequ1}
\widehat \cP_{t-1,t}^{(S^\tau,{\mathbb{G}})}(\xi^{\mathbb{G}})\leq\Xi+ \underset{\delta\in L^0(\cF_{t-1})}\essinf\widehat \cP_{t-1,t}^{(\overline{S},{\mathbb{F}})}(\xi^{\mathbb{F}} +\delta{I}_{\{\widetilde{G}_t=0\}})I_{\{\tau\ge t\}}.
\end{equation} On the other hand,  again due to (\ref{Equlity4Sets}), for any $x^{\mathbb{G}}\in \cP_{t-1,t}^{(S^\tau,{\mathbb{G}})}(\xi^{\mathbb{G}} )$, we deduce the existence of $(x_{t-1},\delta)\in\cP_{t-1,t}^{(\overline{S},{\mathbb{F}})}(\xi^{\mathbb{F}} +\delta{I}_{\{\widetilde{G}_t=0\}})\times{L}^0(\cF_{t-1})$, such that $P$-a.s.,
\begin{equation*}
\begin{split}
x^{\mathbb{G}}&\geq \Xi+x_{t-1}I_{\{\tau\geq{t}\}}\geq \Xi+\widehat\cP_{t-1,t}^{(\overline{S},{\mathbb{F}})}(\xi^{\mathbb{F}} +\delta{I}_{\{\widetilde{G}_t=0\}})I_{\{\tau\geq{t}\}}\\
&\geq \Xi+\underset{\delta\in L^0(\cF_{t-1})}\essinf\widehat \cP_{t-1,t}^{(\overline{S},{\mathbb{F}})}(\xi^{\mathbb{F}}  +\delta{I}_{\{\widetilde{G}_t=0\}})I_{\{\tau\ge t\}}.\end{split}\end{equation*} 
As a result, this implies that 
\begin{equation*} 
\widehat \cP_{t-1,t}^{(S^\tau,{\mathbb{G}})}(\xi)\displaystyle\geq \Xi+I_{\{\tau\ge t\}}\underset{\delta\in L^0(\cF_{t-1})}\essinf\widehat\cP_{t-1,t}^{(\overline{S},{\mathbb{F}})}(\widehat g_t+\delta{I}_{\{\widetilde{G}_t=0\}}),\end{equation*}
and by combining this  with (\ref{Inequ1}), we get (\ref{FirstEquality}). This ends the first part.\\
{\bf Part 2.} This part proves that, for $(\Xi, \xi^{\mathbb{G}},\xi^{\mathbb{F}})\in{L}^0({\cal{G}}_{t-1})\times{L}^0({\cal{G}}_{t})\times{L}^0({\cal{F}}_{t})$,
\begin{equation}\label{Claim}
\begin{split}
&\mbox{If}\quad \cP_{t-1,t}^{(S^\tau,{\mathbb{G}})}(\xi^{\mathbb{G}})=\Xi+L^0_+({\cal{G}}_{t})+\cP_{t-1,t}^{(\widetilde{S},{\mathbb{F}},\widetilde{Q})}(\xi^{\mathbb{F}})I_{\{\tau\geq t\}},\quad\mbox{then}\\ 
&\mbox{we have}\quad \widehat{\cP}_{t-1,t}^{(S^\tau,{\mathbb{G}})}(\xi^{\mathbb{G}})=\Xi+\widehat{\cP}_{t-1,t}^{(\widetilde{S},{\mathbb{F}},\widetilde{Q})}(\xi^{\mathbb{F}})I_{\{\tau\geq t\}}.\end{split}
\end{equation}
To this end we suppose that the left hand side of this implication holds. On the one hand, we notice
$$
\Xi+\cP_{t-1,t}^{(\widetilde{S},{\mathbb{F}},\widetilde{Q})}(\xi^{\mathbb{F}})I_{\{\tau\geq t\}}\subset \cP_{t-1,t}^{(S^\tau,{\mathbb{G}})}(\xi^{\mathbb{G}}),$$ and hence by taking essential infinimum on both sides we get 
\begin{equation}\label{EssentialInf0}
\widehat{\cP}_{t-1,t}^{(S^\tau,{\mathbb{G}})}(\xi^{\mathbb{G}})\leq\Xi+\widehat{\cP}_{t-1,t}^{(\widetilde{S},{\mathbb{F}},\widetilde{Q})}(\xi^{\mathbb{F}})I_{\{\tau\geq t\}}.
\end{equation}
On the other hand, for any $p^{\mathbb{G}}\in \cP_{t-1,t}^{(S^\tau,{\mathbb{G}})}(\xi^{\mathbb{G}})$, there exists $p^{\mathbb{F}}\in \cP_{t-1,t}^{(\widetilde{S},{\mathbb{F}},\widetilde{Q})}(\xi^{\mathbb{F}})$ such that 
$$
p^{\mathbb{G}}\geq \Xi+p^{\mathbb{F}}I_{\{\tau\geq t\}}\geq \Xi+\widehat{\cP}_{t-1,t}^{(\widetilde{S},{\mathbb{F}},\widetilde{Q})}(\xi^{\mathbb{F}})I_{\{\tau\geq t\}},\quad P\mbox{-a.s.}.$$
Therefore, by taking essential infimum in the above inequality and combining the resulting inequality with (\ref{EssentialInf0}), the claim (\ref{Claim}) follows immediately.This ends the second part.\\
{\bf Part 3.}  Hereto, we summarize the proof of the theorem. In fact, in order to prove assertion (a), we appeal to Theorem \ref{TheoremSH1-SetOfPrices}-(a). Then the proof of the first equality in  (\ref{Price-Claim1})  is a direct combination of the first equality in (\ref{Set4PricesClaim1}) and Part 1 applied to $(\xi^{\mathbb{G}},\Xi,\xi^{\mathbb{F}})=(g_t I_{\{\tau>{t}\}},0,\widehat{g}_t)$. The second equality in (\ref{Price-Claim1}) follows from combining the second equality in  (\ref{Set4PricesClaim1}) and Part 2 applied to $(\xi^{\mathbb{G}},\Xi,\xi^{\mathbb{F}})=(g_t I_{\{\tau>{t}\}},0,\widehat{g}_t)$. Similarly, the first  (respectively the second) equality in (\ref{Price-Claim2}) is a direct consequence of the first (respectively the second) equality in (\ref{Set4PricesClaim2}) and Part 1 (respectively Part 2) applied to $(\xi^{\mathbb{G}},\Xi,\xi^{\mathbb{F}})=(g_tI_{\{\tau\geq{t}\}},0,\overline{g}_t)$. Assertion (c) follows from combining  (\ref{Set4PricesClaim3}) and Part 1 and 2 applied  $(\xi^{\mathbb{G}},\Xi,\xi^{\mathbb{F}})=(K_{\tau}I_{\{\tau\leq{t}\}},K_{\tau}I_{\{\tau\leq{t-1}\}},\kappa^{(0)}_t)$. Finally, assertion (d) is direct consequence from combining  (\ref{Set4PricesClaim4}) and Parts 1 and 2 applied to $(\xi^{\mathbb{G}},\Xi,\xi^{\mathbb{F}})=(g_tI_{\{\tau>{t}\}}+K_{\tau}I_{\{\tau\leq{t}\}},K_{\tau}I_{\{\tau\leq{t-1}\}},\kappa^{(g)}_t)$. This completes the proof of theorem.
\end{proof}
\subsection{The general pricing formulas and the prices' dynamics}
Hereto, we fully describe the pricing formulas for the three kind of vulnerable claims, and afterwards we single out the various risks in their dynamics as well. To this end, throughout the rest of the paper, for any two processes $X$ and $Y$, we denote by $X\is Y$ --the stochastic integral of $X$ with respect to $Y$--, $[X,Y]$, and $\langle{X},Y\rangle$ --when it exists-- the processes given by 
\begin{equation}\label{SI-Braccket}
\begin{split}
&X\is Y:=\sum_{s=1}^{\cdot} X_s\Delta{Y}_s:=\sum_{s=1}^t X_s(Y_s-Y_{s-1}),\quad (X\is Y)_0=0,\\
&[X,Y]:=\sum_{s=1}^{\cdot}\Delta{X}_s\Delta{Y}_s,\quad \langle{X},Y\rangle:=\sum_{s=1}^{\cdot} E\left[\Delta{X}_s\Delta{Y}_s\Big|{\cal{F}}_{s-1}\right],\end{split}\end{equation}
where the convention $\sum_{\emptyset}=0$ will be used throughout the paper. For further notation and definitions about stochastic calculus in discrete-time, we refer the reader to  \cite{JS} and  \cite{Yan}. 

 We recall the triplet $(m,N^{\mathbb{G}},D^{o,\mathbb{F}})$ associated to $\tau$, which plays central roles in quantifying the various risks generated by $\tau$, as follows,
\begin{equation}\label{mN(G)}
\begin{split}
m_t:=&1+\sum_{s=1}^t\left(\widetilde{G}_s-E[\widetilde{G}_s\big|{\cal{F}}_{s-1}]\right),\ {N}^{\mathbb{G}}_t:=I_{\{\tau\leq{t}\}}-\sum_{s=1}^{t\wedge\tau}{{P(\tau=s\big|{\cal{F}}_s)}\over{\widetilde{G}_s}},\\
  D^{o,\mathbb{F}}_t&:=\sum_{s=0}^t P(\tau=s\big|{\cal{F}}_s).\end{split}
\end{equation}
The process $m$ is an $\mathbb{F}$-martingale which is a BMO martingale, see \cite{Jeulin} for more details about this fact. Herein, $m$ quantifies the correlation risk resulting from the interaction of $\tau$ with the flow $\mathbb{F}$, see \cite{ChoulliDavelooseVanmaele,ChoulliYansori1,ChoulliYansori2,ChoulliYansori3} for more about this. The process $N^{\mathbb{G}}$ is a $\mathbb{G}$-martingale which was introduced in \cite{ChoulliDavelooseVanmaele}  and called the  main generator of pure-default-martingales of type 1. This process $N^{\mathbb{G}}$ quantifies the main generator of the pure-default-risk borne in the random time, see \cite{ChoulliDavelooseVanmaele,ChoulliDavelooseVanmaeleMortality,ChoulliYansori2,ChoulliYansori3} for more details and related works and results. The process $D^{o,\mathbb{F}}$ is an nondecreasing process $\mathbb{F}$-adapted and is the $\mathbb{F}$-optional dual projection of $D:=1_{\Rbrack\tau,\infty\Rbrack}$. We refer the reader to \cite{ChoulliDavelooseVanmaeleMortality}, for the role of $D^{o,\mathbb{F}}$ in quantifying correlation risks resulting from the interplay between  $\tau$ and the benefit policy in life insurance. The risks in $\mathbb{G}$, which are mainly coming from $\mathbb{F}$, are represented by a transform operator ${\cal{T}}$, defined for any process $M$ by 
\begin{equation}\label{Transform}
{\cal{T}}(M):=\sum_{u=1}^{\tau\wedge\cdot}{{G_{u-1}}\over{\widetilde{G}_u}}(M_u-M_{u-1})+\sum_{u=1}^{\tau\wedge\cdot}E[{I}_{\{\widetilde{G}_u=0\}}(M_u-M_{u-1})\big|{\cal{F}}_{u-1}].
\end{equation}
Herein, if $M$ is an $\mathbb{F}$-martingale, the ${\cal{T}}(M)$ is a $\mathbb{G}$-martingale. For more details about these and related properties, we refer the reader to \cite[Theorem 2.14]{ChoulliDavelooseVanmaele}  and the references therein. \\
Throughout the remaining part of the paper, we use the one-period-pricing operator $\widehat{\cP}^{(\widetilde{S},\widetilde{Q})}_{t,t+1}(\cdot)$, given by Definition \ref{Definition4MinimumPrice}, which we recall below
\begin{equation}\label{1-period-PricingOprator}
\widehat{\cP}^{(\widetilde{S},\widetilde{Q})}_{t,t+1}(\cdot):=\underset{\theta\in{L}^0({\cal{F}}_{t})}{\overset{\widetilde{Q}}{\essinf}}\underset{{{\cal{F}}_t}}{\overset{\widetilde{Q}}{\esssup}}(\theta\Delta\widetilde{S}_{t+1}+\cdot),
\end{equation}
and we will use the following notational abreviation
\begin{equation}\label{RiskTerminolgies}\begin{split}
&\mbox{``CRRisk}(Y_1,Y_2)":=\mbox{Correlation Risk from}\ (Y_1,Y_2),\\
& \mbox{ and ``PFRisk}":= \mbox{Pure Financial Risk}.\end{split}\end{equation}
Furthermore, for any {\it recovery process} $R$, we introduce an important functional $f_R(t, x)=f_R(t,\omega, x)$, which is ${\cal{B}}(\RR)\otimes{\cal{O}}(\mathbb{F})$-measurable, and is given by  
\begin{equation}\label{f(R)4RecopveryR}
f_R(t, x):=xI_{\{\widetilde{G}_t=G_t>0\}}+R_t{I}_{\{\widetilde{G}_t>G_t=0\}}+\max(x,R_t)I_{\{\widetilde{G}_t>G_t>0\}},
\end{equation}
for any $(t,x)\in\{0,..., T,\}\times\RR$.
\begin{theorem}\label{GPriceProcessDynamics-Class1}
 Suppose $(\widetilde{S},\mathbb{F},\widetilde{Q})$ fulfills AIP. Consider $g_T\in  L^1({\cal{F}}_T)$ and the pair of  vulnerable claim  and its associated $\mathbb{F}$-claim $(\xi^{(\mathbb{G},1)},\xi^{(\mathbb{F},1)})$ which belongs to $ \{(g_TI_{\{\tau>T\}},g_TI_{\{G_T>0\}}), (g_TI_{\{\tau\ge T\}}, g_TI_{\{\widetilde{G}_{T}>0\}})\}$. Let $\widehat{\cP}^{(\mathbb{G},1)}$ be the super-hedging price process for $\xi^{(\mathbb{G},1)}$ under $(S^{\tau},\mathbb{G},P)$, and $f_0$ be the functional defined in (\ref{f(R)4RecopveryR}) for the zero-recovery. Then the following assertions hold.\\
{\rm{(a)}} The price process $\widehat{\cP}^{(\mathbb{G},1)}$ is given by  $\widehat{\cP}^{(\mathbb{G},1)}=\widehat{\cP}^{(\mathbb{F},1)}{I}_{\Rbrack0,\tau\Rbrack}$ and
\begin{equation}\label{PriceDynamicsDescription}
\begin{split}
  \widehat{\cP}_t^{(\mathbb{F},1)}=\begin{cases}{\widehat{\cP}}^{(\widetilde{S},\widetilde{Q})}_{t,t+1}\left( f_0(t+1,\widehat{\cP}_{t+1}^{(\mathbb{F},1)})\right),\ \mbox{if}\ {t}< T,\\
\\
\xi^{(\mathbb{F},1)},\hskip 3cm \mbox{if}\  t=T.\end{cases}\end{split}
\end{equation}
{\rm{(b)}} The price process $\widehat{\cP}^{(\mathbb{G},1)}$ can be decomposed into
\begin{equation}\label{GPriceProcessDynamicsEquation}
\begin{split}
\widehat{\cP}^{(\mathbb{G},1)}&=\underbrace{ \widehat{\cP}_0^{(\mathbb{F},1)}{I}_{\{\tau>0\}}+\sum_{s=1}^{\tau\wedge\cdot}E\left[{{G_s}\over{G_{s-1}}} \widehat{\cP}_s^{(\mathbb{F},1)} - \widehat{\cP}_{s-1}^{(\mathbb{F},1)}\big|{\cal{F}}_{s-1}\right]}_{\mbox{super-hedging price's trend}}+\underbrace{{\cal{T}}(M^{(1)})}_{\mbox{PFRisk}}\\
&-\underbrace{ \widehat{\cP}^{(\mathbb{F},1)} \is{N}^{\mathbb{G}}}_{\mbox{Pure-Default-Risk}}-\underbrace{G_{-}^{-1}\is{\cal{T}}(N^{(1)})}_{\mbox{CRRisk}(\tau,\mbox{benefit policy})}\\
&+\underbrace{G_{-}^{-1}\is{\cal{T}}(\overline{N}^{(1)})-G_{-}^{-2}(\Delta\widetilde{V}^{(1)}+\Delta\langle{M}^{(1)},m\rangle)\is{\cal{T}}({m})}_{\mbox{CRRisk}(\tau,\mathbb{F})}.\end{split}\end{equation}
Here  $G_{-}:=G_{\cdot -1}$ and the quadruplet $(M^{(1)}, N^{(1)}, \overline{N}^{(1)}, \widetilde{V}^{(1)})$ is given by 
\begin{equation}\label{(N,Vtilde,M)}\begin{split}
&\widetilde{V}^{(1)}:= \sum_{s=1}^{\cdot} E\left[ \widehat{\cP}_s^{(\mathbb{F},1)} (\widetilde{G}_s-G_s)\big|{\cal{F}}_{s-1}\right],\quad N^{(1)}:= \widehat{\cP}^{(\mathbb{F},1)} \is{D}^{o,\mathbb{F}}- \widetilde{V}^{(1)},\\
&\overline{N}^{(1)}:=\sum_{s=1}^{\cdot} \left( \widehat{\cP}_s^{(\mathbb{F},1)} \widetilde{G}_s-E[ \widehat{\cP}_s^{(\mathbb{F},1)}\widetilde{G}_s\big|{\cal{F}}_{s-1}]\right) -G_{-}\is M^{(1)}-\xi^{(1)}\is{m},\\
& M^{(1)}:=\sum_{s=1}^{\cdot}\left( \widehat{\cP}_s^{(\mathbb{F},1)} -E[ \widehat{\cP}_s^{(\mathbb{F},1)} \big|{\cal{F}}_{s-1}]\right),\quad\xi^{(1)}_s:=E\left[ \widehat{\cP}_s^{(\mathbb{F},1)} \big|{\cal{F}}_{s-1}\right].
\end{split}\end{equation}
\end{theorem}
The theorem claims that the super-hedging price process of the vulnerable claim with payoff $\xi^{(\mathbb{G},1)}$, which does not involve payment after the occurrence of $\tau$, is given by the super-hedging price process  under the model $(\widetilde{S},\mathbb{F},\widetilde{Q})$ for a corresponding $\mathbb{F}$-claim  The theorem presents two cases of this class of vulnerable claims depending whether there is payment at $\tau$ or not. These two cases differ slightly in  the vulnerable claim's payoff and hence in its corresponding $\mathbb{F}$-claim and subsequently in the super-hedging price process of this latter claim. However, both $\mathbb{G}$-super-hedging prices of these two cases have the same structures before $\tau$, and the same risk decomposition structure. The decomposition (\ref{GPriceProcessDynamicsEquation}) of the dynamics' of the price process of the vulnerable claim is important in the securitization and/or hedging as it singles out the various risks and their origins. \\

Below, we discuss some particular cases of the pair $(\tau,\mathbb{F})$ which are frequently addressed in the  credit risk and life insurance literatures.
\begin{remark}
{\rm{(a)}} The case of immersion, which is equivalent to  the martingale $m$ being constant (i.e. $m_t=m_0$ $P$-a.s. for any $t\in\{1,...,T\}$). If immersion holds, then 
\begin{equation}\label{GPriceProcessDynamicsEquation-Immersion}
\begin{split}
\widehat{\cP}^{(\mathbb{G},1)}&=\underbrace{\widetilde{\cP}^{(1)}_0{I}_{\{\tau>0\}}+\sum_{s=1}^{\tau\wedge\cdot}E\left[{{G_s}\over{G_{s-1}}} \widehat{\cP}_s^{(\mathbb{F},1)}  - \widehat{\cP}_{s-1}^{(\mathbb{F},1)} \big|{\cal{F}}_{s-1}\right]}_{\mbox{super-hedging-price's trend}}+\underbrace{{\cal{T}}(M^{(1)})}_{\mbox{PFRisk}}\\
&-\underbrace{ \widehat{\cP}^{(\mathbb{F},1)} \is{N}^{\mathbb{G}}}_{\mbox{Pure-Default-Risk}}-\underbrace{G_{-}^{-1}\is{\cal{T}}(N^{(1)})}_{\mbox{CRRisk}(\tau,\mbox{benefit policy})}.\end{split}\end{equation}
{\rm{(b)}} The case of independence between $\tau$ and $\mathbb{F}$. This is a particular case of immersion, and furthermore in this case we have 
\begin{equation}\label{GPriceProcessDynamicsEquation-Immersion}
\begin{split}
\widehat{\cP}^{(\mathbb{G},1)}&=\underbrace{ \widehat{\cP}_0^{(\mathbb{F},1)}{I}_{\{\tau>0\}}+\sum_{s=1}^{\tau\wedge\cdot}\left({{G_s}\over{G_{s-1}}}E\left[ \widehat{\cP}_s^{(\mathbb{F},1)} \big|{\cal{F}}_{s-1}\right] - \widehat{\cP}_{s-1}^{(\mathbb{F},1)} \right)}_{\mbox{super-hedging-price's trend}}\\
&+{{G}\over{G_{-}}}\is{\cal{T}}(M^{(1)})-\underbrace{ \widehat{\cP}^{(\mathbb{F},1)} \is{N}^{\mathbb{G}}}_{\mbox{Pure-Default-Risk}}.\end{split}\end{equation}
\end{remark}
The statements in the remark follows directly from Theorem \ref{GPriceProcessDynamics-Class1}, and hence their proofs will be omitted herein. The second main result of this subsection deals with vulnerable claims that involve payment at $\tau$, and these claims take two forms depending whether there is payment before the occurrence of $\tau$ or not. In life insurance, these vulnerable claims are known as policies where the beneficiaries receive payments at the moment of death, and for which there are two types of policies depending whether there is benefit upon survival or not. Below, we start with the case where there is no benefit upon survival.
\begin{theorem}\label{GPriceProcessDynamics-Class2(g=0)} Let the recovery process $K$ be an $\mathbb{F}$-adapted and integrable process, $f_K$ given by (\ref{f(R)4RecopveryR}),  $\xi^{\mathbb{G},2}=K_{\tau}{I}_{\{\tau\leq{T}\}}$ be the vulnerable claim's payoff, and $\widehat{\cP}^{(\widetilde{S},\widetilde{Q})}_{t,t+1}(\cdot)$ be the one-period-pricing operator given by (\ref{1-period-PricingOprator}). Suppose that $(\widetilde{S},\mathbb{F},\widetilde{Q})$ fulfills AIP (equivalently $(S^{\tau},\mathbb{G})$ fulfills AIP ) and denote by $\widehat{\cP}^{(\mathbb{G},2)}$ the super-hedging price process for the claim $\xi^{\mathbb{G},2}$. Then the following assertions hold.\\
 {\rm{(a)}} The process $\widehat{\cP}^{(\mathbb{G},2)}$ is given by $\widehat{\cP}^{(\mathbb{G},2)}=K_{\tau}I_{\Rbrack\tau,\infty\Rbrack}+ \widehat{\cP}^{(\mathbb{F},2)} {I}_{\Rbrack0,\tau\Rbrack}$ and
 \begin{equation}\label{PriceDynamicsDescription1}
  \widehat{\cP}_t^{(\mathbb{F},2)} =\begin{cases}\widehat{\cP}^{(\widetilde{S},\widetilde{Q})}_{t,t+1}\bigl(f_K( t+1, \widehat{\cP}_{t+1}^{(\mathbb{F},2)})\bigr),\quad\mbox{if}\quad t<T\\
  0\hskip 4cm\mbox{if}\quad t=T\end{cases}.
\end{equation}
{\rm{(b)}} The dynamics of $\widehat{\cP}^{(\mathbb{G},2)}$ can be decomposed 
\begin{equation}\label{Dynamics4P(2,G)}
\begin{split}
\widehat{\cP}^{(\mathbb{G},2)}=&K_0{I}_{\{\tau=0\}}+\widehat{\cP}_0^{(\mathbb{F},2)} {I}_{\{\tau>0\}}+\underbrace{{\cal{T}}(M^{(2)})}_{\mbox{PFRisk}}+\underbrace{(K-\widehat{\cP}^{(\mathbb{F},2)})\is{N}^{\mathbb{G}}}_{\mbox{Pure-Default-Risk}}\\
&+\underbrace{G_{-}^{-1}\is{\cal{T}}(N^{(2)})}_{\mbox{CRRisk}(\tau,\mbox{benefit policy})}\\
&+\underbrace{G_{-}^{-1}\is{\cal{T}}(\overline{N}^{(2)})-G_{-}^{-2}(\Delta\widetilde{V}^{(2)}+\Delta\langle{M}^{(2)},m\rangle)\is{\cal{T}}({m})}_{\mbox{CRRisk}(\tau,\mathbb{F})}\\
&+\underbrace{\sum_{s=1}^{\tau\wedge\cdot}E\left[K_s{{\widetilde{G}_s-G_s}\over{G_{s-1}}}+\widehat{\cP}_s^{(\mathbb{F},2)} {{G_s}\over{G_{s-1}}}-\widehat{\cP}_{s-1}^{(\mathbb{F},2)}\Big|{\cal{F}}_{s-1}\right]}_{\mbox{super-hedging-price's trend}}.
\end{split}\end{equation}
Here, the quadruplet $(M^{(2)},A^{(2)}, N^{(2)}, \widetilde{V}^{(2)})$ is given by 
\begin{equation}\label{(N,Vtilde,M)}\begin{split}
N^{(2)}:&=(K-\widehat{\cP}^{(\mathbb{F},2)})\is{D}^{o,\mathbb{F}}- \widetilde{V}^{(2)},\\
\widetilde{V}^{(2)}&:=\sum_{s=1}^{\cdot}\Delta\widetilde{V}^{(2)}_s:=\sum_{s=1}^{\cdot}E\left[(K_s-\widehat{\cP}_s^{(\mathbb{F},2)})(\widetilde{G}_s-G_s)\big|{\cal{F}}_{s-1}\right]\\
\overline{N}^{(2)}&:=\sum_{s=1}^{\cdot} \left(\widehat{\cP}_s^{(\mathbb{F},2)}\widetilde{G}_s-E[\widehat{\cP}_s^{(\mathbb{F},2)}\widetilde{G}_s\big|{\cal{F}}_{s-1}]\right) -G_{-}\is M^{(2)}-\xi^{(2)}\is{m},\\
M_t^{(2)}&:=\sum_{s=1}^t\left(\widehat{\cP}_s^{(\mathbb{F},2)}-E[\widehat{\cP}_s^{(\mathbb{F},2)}\big|{\cal{F}}_{s-1}]\right),\quad\xi^{(2)}_t:=E\left[\widehat{\cP}_t^{(\mathbb{F},2)}\big|{\cal{F}}_{t-1}\right].
\end{split}\end{equation}
\end{theorem}
Below, we treat the case where there are both benefits, upon survival and at the moment of death $\tau$.
\begin{theorem}\label{GPriceProcessDynamics-Class2}Let $g_T\in  L^1({\cal{F}}_T)$, the recovery process $K$ be integrable and $\mathbb{F}$-adapted, $f_K$ be the functional given by  (\ref{f(R)4RecopveryR}), $\xi^{(\mathbb{G},3)}:=g_TI_{\{\tau>T\}}+K_{\tau}{I}_{\{\tau\leq{T}\}}$ be the vulnerable claim's payoff. If $(\widetilde{S},\mathbb{F},\widetilde{Q})$ fulfills AIP and   $\widehat{\cP}^{(\mathbb{G},3)}$ denotes the super-hedging price for $\xi^{(\mathbb{G},3)}$, then the following assertions hold.\\
{\rm{(a)}} The process $\widehat{\cP}^{(\mathbb{G},3)}$ is given by $\widehat{\cP}^{(\mathbb{G},3)}=K_{\tau}I_{\Rbrack\tau,\infty\Rbrack}+\widehat{\cP}^{(\mathbb{F},3)}{I}_{\Rbrack0,\tau\Rbrack}$ and 
 \begin{equation}\label{PriceDynamicsDescription(g)}
\begin{split}
\ \widehat{\cP}_t^{(\mathbb{F},3)}=\begin{cases}\widehat{\cP}^{(\widetilde{S},\widetilde{Q})}_{t,t+1}\bigl(f_K(t+1, \widehat{\cP}_{t+1}^{(\mathbb{F},3)})\bigr),\ \mbox{if}\quad t<T,\\
\\
 g_T{I}_{\{G_T>0\}},\hskip 2.4cm \mbox{if}\ t=T.\end{cases}\end{split}
\end{equation}
{\rm{(b)}} The dynamics of $\widehat{\cP}^{(\mathbb{G},3)}$ can be decomposed 
\begin{equation}\label{Dynamics4P(3,G)}
\begin{split}
\widehat{\cP}^{(\mathbb{G},3)}&=K_0{I}_{\{\tau=0\}}+\widehat{\cP}_0^{(\mathbb{F},3)}{I}_{\{\tau>0\}}+\underbrace{(K-\widehat{\cP}^{(\mathbb{F},3)})\is{N}^{\mathbb{G}}}_{\mbox{Pure-Default-Risk}}+\underbrace{{\cal{T}}(M^{(3)})}_{\mbox{PFRisk}}\\
&+\underbrace{{1\over{G_{-}}}\is{\cal{T}}(N^{(3)})}_{\mbox{CRRisk}(\tau,\mbox{benefit policy})}+\underbrace{{1\over{G_{-}}}\is{\cal{T}}(\overline{N}^{(3)})-{{(\Delta\widetilde{V}^{(3)}+\Delta\langle{M}^{(3)},m\rangle)}\over{G_{-}^2}}\is{\cal{T}}({m})}_{\mbox{CRRisk}(\tau,\mathbb{F})}\\
&+\underbrace{\sum_{s=1}^{\tau\wedge\cdot}E\left[K_s{{\widetilde{G}_s-G_s}\over{G_{s-1}}}+\widehat{\cP}_s^{(\mathbb{F},3)}{{G_s}\over{G_{s-1}}}-\widehat{\cP}_{s-1}^{(\mathbb{F},3)}\Big|{\cal{F}}_{s-1}\right]}_{\mbox{super-hedging price's trend}}
\end{split}\end{equation}
Here $(M^{(3)}, N^{(3)}, \overline{N}^{(3)}, \widetilde{V}^{(3)})$ are given by 
\begin{equation}\label{(N,Vtilde,M)}\begin{split}
N^{(3)}:&=(K-\widehat{\cP}^{(\mathbb{F},3)})\is{D}^{o,\mathbb{F}}- \widetilde{V}^{(3)},\\
\widetilde{V}^{(3)}&:=\sum_{s=1}^{\cdot}\Delta\widetilde{V}^{(3)}_s:=\sum_{s=1}^{\cdot}E\left[(K_s-\widehat{\cP}_s^{(\mathbb{F},3)})(\widetilde{G}_s-G_s)\big|{\cal{F}}_{s-1}\right]\\
\overline{N}^{(3)}&:=\sum_{s=1}^{\cdot} \left(\widehat{\cP}_s^{(\mathbb{F},3)}\widetilde{G}_s-E[\widehat{\cP}_s^{(\mathbb{F},3)}\widetilde{G}_s\big|{\cal{F}}_{s-1}]\right) -G_{-}\is M^{(3)}-\xi^{(3)}\is{m},\\
M_t^{(3)}&:=\sum_{s=1}^t\left(\widehat{\cP}_s^{(\mathbb{F},3)}-E[\widehat{\cP}_s^{(\mathbb{F},3)}\big|{\cal{F}}_{s-1}]\right),\quad\xi^{(3)}_s:=E\left[\widehat{\cP}_s^{(\mathbb{F},3)}\big|{\cal{F}}_{s-1}\right],
\end{split}\end{equation}
\end{theorem}
We end this subsection by illustrating the main results of Theorems \ref{GPriceProcessDynamics-Class1}, \ref{GPriceProcessDynamics-Class2(g=0)} and \ref{GPriceProcessDynamics-Class2} on the case of {\it vulnerable options} with and/or without recovery. 
\begin{theorem}\label{Results4Options} Suppose that $g_T\in  L^0_+({\cal{F}}_T)$, and the recovery process $K$ be  nonnegative integrable and $\mathbb{F}$-adapted. Let $\xi^{(\mathbb{F},1)}\in\{g_TI_{\{G_T>0\}},g_TI_{\{\widetilde{G}_T>0\}}\}$ and put $\overline{K}:=KI_{\{\widetilde{G}>G\}}$. Then the three processes $ \widehat{\cP}^{(\mathbb{F},i)}$, $i=1,2,3$, given in Theorems \ref{GPriceProcessDynamics-Class1}, \ref{GPriceProcessDynamics-Class2(g=0)} and \ref{GPriceProcessDynamics-Class2} respectively, satisfy the following

\begin{equation}\label{EquationsPhat(1,2,3)}
\begin{split}
&\widehat{\cP}_t^{(\mathbb{F},1)}=\begin{cases}{\widehat{\cP}}^{(\widetilde{S},\widetilde{Q})}_{t,t+1}\left( \widehat{\cP}_{t+1}^{(\mathbb{F},1)}\right),\ \mbox{if}\ {t}\leq T-1,\\ \xi^{(\mathbb{F},1)},\hskip 1.7cm\mbox{if}\  t=T\end{cases},\\
&  \widehat{\cP}_t^{(\mathbb{F},2)} =\begin{cases}\widehat{\cP}^{(\widetilde{S},\widetilde{Q})}_{t,t+1}\bigl(\max(  \widehat{\cP}_{t+1}^{(\mathbb{F},2)} ,\overline{K}_{t+1})\bigr),\ \mbox{if}\  t\leq T-1,\\
\\
  \widehat{\cP}_T^{(\mathbb{F},2)} =0,\quad\quad\quad\quad\quad\quad\quad\quad\mbox{if}\  t=T\end{cases},\\
 &\mbox{and}\quad \widehat{\cP}_t^{(\mathbb{F},3)}=\begin{cases}\widehat{\cP}^{(\widetilde{S},\widetilde{Q})}_{t,t+1}\bigl(\max( \widehat{\cP}_{t+1}^{(\mathbb{F},3)},\overline{K}_{t+1})\bigr),\ \mbox{if}\ t\leq T-1,\\
\\
 \widehat{\cP}_T^{(\mathbb{F},3)}=g_T{I}_{\{G_T>0\}},\quad\quad\quad\quad\mbox{if}\ t=T\end{cases}.\end{split}
 \end{equation}
\end{theorem}
\begin{proof} For any nonnegative process $R$ and any $x\in\RR^+$,  consider $f_R(t,x)$ defined in (\ref{f(R)4RecopveryR}) and derive
\begin{equation}\label{Functionf(R,x)}
\begin{split}
&f_R(t, x)\\
&:=xI_{\{\widetilde{G}_t=G_t>0\}}+R_t{I}_{\{\widetilde{G}_t>G_t=0\}}+\max(x,R_t)I_{\{\widetilde{G}_t>G_t>0\}}\\
&=\max(xI_{\{G_t>0\}}, R_t{I}_{\{\widetilde{G}_t>G_t\}})\left(I_{\{\widetilde{G}_t=G_t>0\}}+{I}_{\{\widetilde{G}_t>G_t=0\}}+I_{\{\widetilde{G}_t>G_t>0\}}\right)\\
&=\max(xI_{\{G_t>0\}}, R_t{I}_{\{\widetilde{G}_t>G_t\}}).
\end{split}
\end{equation}
 Therefore, the proof of the three equalities in (\ref{EquationsPhat(1,2,3)}) follow immediately from combining this last equality in (\ref{Functionf(R,x)}), and assertion (a) for Theorems \ref{GPriceProcessDynamics-Class1}, \ref{GPriceProcessDynamics-Class2(g=0)} and \ref{GPriceProcessDynamics-Class2} respectively. This ends the proof of the theorem.
\end{proof}

\begin{remark} Under the assumptions of Theorem \ref{Results4Options}, the following hold
\begin{equation}\label{ComparingP(1)andP(3)}
\widehat{\cP}^{(\mathbb{G},3)}\geq\max\left( \widehat{\cP}^{(\mathbb{G},1)},\widehat{\cP}^{(\mathbb{G},2)}\right)\quad\mbox{and}\quad \widehat{\cP}^{(\mathbb{F},3)}\geq \max\left(\widehat{\cP}^{(\mathbb{F},1)},\widehat{\cP}^{(\mathbb{F},2)}\right).
\end{equation}
\end{remark}
\subsection{Proofs of Theorems \ref{GPriceProcessDynamics-Class1}, \ref{GPriceProcessDynamics-Class2(g=0)} and \ref{GPriceProcessDynamics-Class2}}
These proofs rely on the following three technical lemmas, which are interesting in themselves.
\begin{lemma}\label{AIP4Qtilde}
Suppose that $(\widetilde{S},\mathbb{F},\widetilde{Q})$ satisfies the AIP condition, and let $\widehat{\cP}^{(\mathbb{F},3)}$ be defined in (\ref{PriceDynamicsDescription(g)}). Then 
\begin{equation}\label{Properties4Ptilde(3)}
\widehat{\cP}^{(\mathbb{F},3)}\geq 0,\quad\mbox{and}\quad \widehat{\cP}^{(\mathbb{F},3)}I_{\{G=0\}}\equiv 0.
\end{equation}
\end{lemma}
The proof of this lemma will be omitted herein. 
\begin{lemma}\label{Lemma4SomeGmartingales} For any $\mathbb{F}$-martingale $M$ and $\mathbb{F}$-predictable process $V$, we have 
\begin{equation}
G_{-}{\widetilde{G}}^{-1}\is{V}^{\tau}={^{p,\mathbb{F}}}(I_{\{\widetilde{G}>0\}})\is{V}^{\tau}-G_{-}^{-1}\Delta{V}\is{\cal{T}}(m),\end{equation}
and 
\begin{equation}
\begin{split}
M^{\tau}=&{\cal{T}}(M)+G_{-}^{-1}\is{\cal{T}}([M,m]-\langle{M},m\rangle)\\
&-G_{-}^{-2}\Delta\langle{M},m\rangle\is{\cal{T}}(m)+G_{-}^{-1}\is\langle{M},m\rangle^{\tau}.
\end{split}\end{equation}
\end{lemma}
The proof of this lemma will be omitted herein. 
\begin{lemma}\label{BSDE4EssentialSup} Let $(X,\mathbb{H},Q)$ be a model defined on the probability space $(\Omega, {\cal{G}},P)$, and satisfying AIP. Consider  $H\in L^0({\cal{H}}_T)$ and a functional $f(t,\omega,x)$ such that for $x\in\R$, $(f(t,x))_{t=0,...,T}$ is $\mathbb{H}$-adapted. Then the following backward stochastic equation
\begin{equation}\label{BSDE}
Y_t={\widehat{\cP}}^{(X,\mathbb{H},Q)}_{t,t+1}(f(t+1,Y_{t+1})),\quad t\leq T-1,\quad Y_T=H,\end{equation}
 has a unique solution. 
\end{lemma}
The proof of this lemma is immediate and will be omitted herein.  The rest of this subsection is devoted to the proof of the three theorems. To this end, and for the sake of simplifying the notation, throughout the proofs we put
$$
 {\widetilde{\cP}}^{(1)}:=\widehat{\cP}^{(\mathbb{F},1)},\quad  {\widetilde{\cP}}^{(2)}:=\widehat{\cP}^{(\mathbb{F},2)},\quad  {\widetilde{\cP}}^{(3)}:=\widehat{\cP}^{(\mathbb{F},3)}.$$
\begin{proof}{\it of Theorems \ref{GPriceProcessDynamics-Class1}, \ref{GPriceProcessDynamics-Class2(g=0)} and \ref{GPriceProcessDynamics-Class2}:} Throughout this proof, we assume that  $(\widetilde{S},\mathbb{F},\widetilde{Q})$ fulfills AIP and consider $g_T\in{L}^1({\cal{F}}_T)$ and an $\mathbb{F}$-adapted and integrable process $K$. The rest of this proof is divided into three parts, where we prove Theorems  \ref{GPriceProcessDynamics-Class1},   \ref{GPriceProcessDynamics-Class2(g=0)} and \ref{GPriceProcessDynamics-Class2} respectively. \\
{\bf Part 1.} This part addresses Theorem \ref{GPriceProcessDynamics-Class1}, and proves it using Theorem \ref{GPriceProcessDynamics-Class2}. \\
1) Suppose that $(\xi^{(\mathbb{G},1)}, \xi^{(\mathbb{F},1)})=(g_T{I}_{\{\tau>T\}},g_T{I}_{\{G_T>0\}})$, and put $K\equiv 0$ in the second equality of (\ref{PriceDynamicsDescription(g)}). Then we deduce that 
\begin{equation}\label{Equality100}
\xi^{(\mathbb{G},3)}=\xi^{(\mathbb{G},1)}\ \mbox{and}\  {\widetilde{\cP}}_t^{(3)}=\begin{cases}\widehat{\cP}^{(\widetilde{S},\widetilde{Q})}_{t,t+1}\bigl(f_0(t+1, {\widetilde{\cP}}_{t+1}^{(3)})\bigr),\ t\leq T-1,\\ {\widetilde{\cP}}_T^{(3)}=g_T{I}_{\{G_T>0\}}.\end{cases}
\end{equation}
Hence, by combining this with Lemma \ref{BSDE4EssentialSup}, we conclude that $\widetilde{\cP}^{(3)}=\widetilde{\cP}^{(1)}$, and hence $\widehat{\cP}^{(\mathbb{G},3)}=\widehat{\cP}^{(\mathbb{G},1)}$ follows immediately as well. This proves assertion (a) for the pair of claims $(g_T{I}_{\{\tau>T\}},g_T{I}_{\{G_T>0\}})$.\\
2) Suppose $(\xi^{(\mathbb{G},1)},\xi^{(\mathbb{F},1)})=(g_TI_{\{\tau\geq{T}\}}, g_T{I}_{\{\widetilde{G}_T>0\}})$.  By combining Lemma \ref{OnePerdiod2MultiPeriod} and Theorem \ref{TheoremSH1-MinimalPrices}-(b), we derive the following
\begin{equation}\label{Equaliy90}
\widehat{\cP}^{(\mathbb{G},3)}_{T-1}={\widehat{\cP}}^{(S^{\tau},\mathbb{G})}_{T-1,T}(\xi^{(\mathbb{G},1)})={\widehat{\cP}}^{(\widetilde{S},\mathbb{F},\widetilde{Q})}_{T-1,T}(g_T{I}_{\{G_T>0\}})I_{\{\tau\geq T\}}.
\end{equation}
 Then, in virtue of  $(\tau>T-1)=(\tau\geq T)$ and $\widetilde{\cP}^{(1)}_T=g_TI_{\{\widetilde{G}_{T}>0\}}$, we obtain 
\begin{equation}\label{Equaliy91}
\widehat{\cP}^{(\mathbb{G},3)}_{T-1}={\widehat{\cP}}^{(\widetilde{S},\mathbb{F},\widetilde{Q})}_{T-1,T}(g_T)I_{\{\tau\geq T\}}={\widehat{\cP}}^{(\widetilde{S},\mathbb{F},\widetilde{Q})}_{T-1,T}(\widetilde{\cP}^{(1)}_T)I_{\{\tau> T-1\}}=\widetilde{\cP}^{(1)}_{T-1}I_{\{\tau> T-1\}}.
\end{equation}
Thus, on the one hand, this latter equality proves (\ref{PriceDynamicsDescription}) for $t=T-1$. On the other hand, as we stated in Remark \ref{Remark4ClaimsOneStepPricing}-(a), this equality tells us that after this one-step we fall into the setting of the first claim, i.e. the case of $g_tI_{\{\tau>t\}}$ with $t=T-1$  and $g_{T-1}=\widetilde{\cP}^{(1)}_{T-1}$ instead. Thus, thanks to step 1 above, we deduce that   (\ref{PriceDynamicsDescription})  holds for any $t\in\{0,...,T-1\}$, and the proof of assertion for the pair of claims $(g_TI_{\{\tau\geq{T}\}}, g_T{I}_{\{\widetilde{G}_T>0\}})$ is complete. \\
3) Thanks to steps 1 and 2 above, we deduce that when $K\equiv 0$ we get $(\widehat{\cP}^{(\mathbb{G},3)},\widetilde{\cP}^{(3)})=(\widehat{\cP}^{(\mathbb{G},1)},\widetilde{\cP}^{(1)})$. Then by combining this with Theorem \ref{GPriceProcessDynamics-Class2}-(b), where we put $K=0$, we conclude that the quadruplets $(M^{(3)}, N^{(3)}, \overline{N}^{(3)}, \widetilde{V}^{(3)})$ and $(M^{(1)}, N^{(1)}, \overline{N}^{(1)}, \widetilde{V}^{(1)})$ coincide, and hence assertion (b) follows immediately. This ends the proof of Theorem \ref{GPriceProcessDynamics-Class1}.\\
{\bf Part 2.} Hereto, we prove Theorem \ref{GPriceProcessDynamics-Class2(g=0)} using again Theorem \ref{GPriceProcessDynamics-Class2}. In fact, in this case we put $g_T=0$ in Theorem \ref{GPriceProcessDynamics-Class2}, and get 
$$\xi^{(\mathbb{G},3)}=\xi^{(\mathbb{G},2)},\quad\mbox{and}\quad {\widetilde{\cP}}_t^{(3)}=\widehat{\cP}^{(\widetilde{S},\widetilde{Q})}_{t,t+1}\bigl(f_K(t+1, {\widetilde{\cP}}_{t+1}^{(3)})\bigr),\quad t\leq T-1,
\quad
 {\widetilde{\cP}}_T^{(3)}=0.$$
Hence, by combining these with  Lemma \ref{BSDE4EssentialSup}, we obtain $ {\widetilde{\cP}}^{(3)}= {\widetilde{\cP}}^{(2)}$ and  $\widehat{\cP}^{(\mathbb{G},3)}=\widehat{\cP}^{(\mathbb{G},2)}$. Therefore, the proof of Theorem \ref{GPriceProcessDynamics-Class2(g=0)}-(a) follows immediately, and both quadruplets   $(M^{(3)}, N^{(3)}, \overline{N}^{(3)}, \widetilde{V}^{(3)})$ and $(M^{(2)}, N^{(2)}, \overline{N}^{(2)}, \widetilde{V}^{(2)})$ coincide. This yields assertion (b) of Theorem \ref{GPriceProcessDynamics-Class2(g=0)} and completes its proof. \\
{\bf Part 3.} This part proves Theorem  \ref{GPriceProcessDynamics-Class2}. To this end, we start proving (\ref{PriceDynamicsDescription(g)}).  By applying Lemma \ref{OnePerdiod2MultiPeriod}  to $(S^{\tau},\mathbb{G})$ for the claim $\xi^{\mathbb{G}}$, and using Theorem \ref{TheoremSH1-MinimalPrices}-(d) afterwards,  we derive 
\begin{equation}\label{Equality110}
\begin{split}
\widehat{\cP}^{(\mathbb{G},3)}_{T-1}&={\widehat{\cP}}^{(S^{\tau},\mathbb{G})}_{T-1,T}(\xi^{\mathbb{G}})=K_{T-1}I_{\{\tau\leq T-1\}}+{\widehat{\cP}}^{(\widetilde{S},\mathbb{F},\widetilde{Q})}_{T-1,T}({\kappa^{(g)}_T})I_{\{\tau> T-1\}}\\
&=K_{T-1}I_{\{\tau\leq T-1\}}+{\widehat{\cP}}^{(\widetilde{S},\mathbb{F},\widetilde{Q})}_{T-1,T}(f_K(T, g_T{I}_{\{G_T>0\}})I_{\{\tau> T-1\}}\\
&=K_{T-1}I_{\{\tau\leq T-1\}}+{\widehat{\cP}}^{(\widetilde{S},\mathbb{F},\widetilde{Q})}_{T-1,T}(f_K(T, \widetilde{\cP}^{(3)}_T)I_{\{\tau> T-1\}}\\
&=K_{T-1}I_{\{\tau\leq T-1\}}+\widetilde{\cP}^{(3)}_{T-1}I_{\{\tau> T-1\}}.\end{split}
\end{equation}
This proves the equality (\ref{PriceDynamicsDescription(g)}) for $t=T-1$. Thus, to prove this equlity for $t=T-2$, we apply again Lemma \ref{OnePerdiod2MultiPeriod} and Theorem \ref{TheoremSH1-MinimalPrices}-(d) afterwards for the pair $(K, \widetilde{\cP}^{(3)})$ instead of $(K,g)$, and write
\begin{equation}\label{Equality112}
\begin{split}
\widehat{\cP}^{(\mathbb{G},3)}_{T-2}&={\widehat{\cP}}^{(S^{\tau},\mathbb{G})}_{T-2,T-1}(\widehat{\cP}^{(\mathbb{G},3)}_{T-1})\\
&=K_{T-2}I_{\{\tau\leq T-2\}}+{\widehat{\cP}}^{(\widetilde{S},\mathbb{F},\widetilde{Q})}_{T-2,T-1}(\kappa_{T-1}(K,\widetilde{\cP}^{(3)}))I_{\{\tau> T-2\}}\\
&=K_{T-2}I_{\{\tau\leq T-2\}}+{\widehat{\cP}}^{(\widetilde{S},\mathbb{F},\widetilde{Q})}_{T-2,T-1}(f_K(T-1,\widetilde{\cP}^{(3)}_{T-1})I_{\{\tau> T-2\}}\\
&=K_{T-2}I_{\{\tau\leq T-2\}}+\widetilde{\cP}^{(3)}_{T-2}I_{\{\tau> T-2\}}
\end{split}
\end{equation}
Therefore, we obtain the equality (\ref{PriceDynamicsDescription(g)}) for $t=T-2$. Thus, we can prove the equality (\ref{PriceDynamicsDescription(g)}) for any $t$ by backward induction. In fact, we suppose that (\ref{PriceDynamicsDescription(g)}) holds for $t+1$, and we will prove it holds for $t$. To this end, we apply then  Lemma \ref{OnePerdiod2MultiPeriod} and Theorem \ref{TheoremSH1-MinimalPrices}-(d) afterwards and get 
\begin{equation}\label{Equality113}
\begin{split}
\widehat{\cP}^{(\mathbb{G},3)}_t&={\widehat{\cP}}^{(S^{\tau},\mathbb{G})}_{t,t+1}(\widehat{\cP}^{(\mathbb{G},3)}_{t+1})=K_{t}I_{\{\tau\leq t\}}+{\widehat{\cP}}^{(\widetilde{S},\mathbb{F},\widetilde{Q})}_{t,t+1}(\kappa_{t+1}(K,\widetilde{\cP}^{(3)}))I_{\{\tau> t\}}\\
&=K_{t}I_{\{\tau\leq t\}}+{\widehat{\cP}}^{(\widetilde{S},\mathbb{F},\widetilde{Q})}_{t,t+1}(f_K(t+1,\widetilde{\cP}^{(3)}_{t+1}))I_{\{\tau> t\}}\\
&=K_{t}I_{\{\tau\leq t\}}+\widetilde{\cP}^{(3)}_{t}I_{\{\tau> t\}}.
\end{split}
\end{equation}
Hence, the equality  (\ref{PriceDynamicsDescription(g)}) holds for $t$, and hence it holds for any $t\in\{0,...,T-1\}$. This completes the proof of (\ref{PriceDynamicsDescription(g)}), while the rest of this part addresses  (\ref{Dynamics4P(3,G)}). By combining (\ref{PriceDynamicsDescription(g)}) and the two facts $XI_{\Rbrack0,\tau\Rbrack}=X^{\tau}-X_{\tau}D$ and $X_{\tau}D=X_{\tau}I_{\Rbrack\tau,\infty\Rbrack}=X_0I_{\{\tau=0\}}+X\is{D}$, which hold for any process $X$, we get $\widehat{\cP}^{(\mathbb{G},3)}_0=K_0I_{\{\tau=0\}}+\widetilde{\cP}^{(3)}_0)I_{\{\tau>0\}}$ and 
\begin{equation}\label{Dynamics1}
\begin{split}
&\widehat{\cP}^{(\mathbb{G},3)}\\
=&(K_0-\widetilde{\cP}^{(3)}_0)I_{\{\tau=0\}}+(K-\widetilde{\cP}^{(3)})\is{D}+(\widetilde{\cP}^{(3)})^{\tau}\\
=&\widehat{\cP}^{(\mathbb{G},3)}_0+(K-\widetilde{\cP}^{(3)})\is{N}^{\mathbb{G}}+{{K-\widetilde{\cP}^{(3)}}\over{\widetilde{G}}}I_{\Rbrack0,\tau\Rbrack}\is{D}^{o,\mathbb{F}}+(\widetilde{\cP}^{(3)})^{\tau}-\widetilde{\cP}^{(3)}_0.
\end{split}
\end{equation}
The last equality is due to $D=N^{\mathbb{G}}+{\widetilde{G}}^{-1}I_{\Lbrack0,\tau\Lbrack}\is{D}^{o,\mathbb{F}}$, see (\ref{mN(G)}) for details. Thanks to (\ref{(N,Vtilde,M)}), it is clear that $M^{(3)}_0=A^{(3)}_0=0$ and 
$${\widetilde{\cP}}^{(3)}={\widetilde{\cP}}^{(3)}_0+M^{(3)}+A^{(3)},\ \mbox{where}\  {A}_t^{(3)}:=\sum_{s=1}^t E[{\widetilde{\cP}}_s^{(3)}-{\widetilde{\cP}}_{s-1}^{(3)}\big|{\cal{F}}_{s-1}],\quad t=1,...,T.$$
Then by inserting these in (\ref{Dynamics1}) and applying Lemma \ref{Lemma4SomeGmartingales}  to $ \widetilde{V}^{(3)}$, $M^{(3)}$ and $N^{(3)}$, we derive 
\begin{equation}\label{Dynamics2}
\begin{split}
&\widehat{\cP}^{(\mathbb{G},3)}=\widehat{\cP}^{(\mathbb{G},3)}_0+(K-\widetilde{\cP}^{(3)})\is{N}^{\mathbb{G}}+{{K-\widetilde{\cP}^{(3)}}\over{\widetilde{G}}}I_{\Rbrack0,\tau\Rbrack}\is{D}^{o,\mathbb{F}}+(\widetilde{\cP}^{(3)})^{\tau}-\widetilde{\cP}^{(3)}_0\\
&=\widehat{\cP}^{(\mathbb{G},3)}_0+(K-\widetilde{\cP}^{(3)})\is{N}^{\mathbb{G}}+(M^{(3)})^{\tau}+(A^{(3)})^{\tau}+{1\over{\widetilde{G}}}\is (N^{(3)}+ \widetilde{V}^{(3)})^{\tau},\\
&=\widehat{\cP}^{(\mathbb{G},3)}_0+(K-\widetilde{\cP}^{(3)})\is{N}^{\mathbb{G}}+{\cal{T}}(M^{(3)})+{1\over{G_{-}}}\is{\cal{T}}(\overline{N}^{(3)})+G_{-}^{-1}\is{\cal{T}}(N^{(3)})\\
&\hskip 0.5cm-{{\Delta\langle{M}^{(3)},m\rangle}\over{G_{-}^2}}\is{\cal{T}}(m)+{1\over{G_{-}}}\is\langle{M}^{(3)},m\rangle^{\tau}-{{\Delta\widetilde{V}^{(3)}}\over{G_{-}^2}}\is{\cal{T}}(m)\\
&\hskip 0.5cm+{{^{p,\mathbb{F}}}\over{G_{-}}}(I_{\{\widetilde{G}>0\}})\is(\widetilde{V}^{(3)})^{\tau}+(A^{(3)})^{\tau}-G_{-}^{-1}I_{\Rbrack0,\tau\Rbrack}\is\left(I_{\{\widetilde{G}=0\}}\is{N}^{(3)}\right)^{p,\mathbb{F}},\\
&=\widehat{\cP}^{(\mathbb{G},3)}_0+(K-\widetilde{\cP}^{(3)})\is{N}^{\mathbb{G}}+{\cal{T}}(M^{(3)})+{1\over{G_{-}}}\is{\cal{T}}(\overline{N}^{(3)})+{1\over{G_{-}}}\is{\cal{T}}(N^{(3)})\\
&-\left(\Delta\langle{M}^{(3)},m\rangle+\Delta\widetilde{V}^{(3)}\right)G_{-}^{-2}\is{\cal{T}}(m)\\
&+G_{-}^{-1}\is(\widetilde{V}^{(3)})^{\tau}+G_{-}^{-1}\is\langle{M}^{(3)},m\rangle^{\tau}+(A^{(3)})^{\tau}.
\end{split}
\end{equation}
The last equality is due to $I_{\{\widetilde{G}=0\}}\is D^{o,\mathbb{F}}=0$. Thus, direct calculations yields
\begin{equation*}
\begin{split}
&G_{-}^{-1}\is(\widetilde{V}^{(3)})^{\tau}+G_{-}^{-1}\is\langle{M}^{(3)},m\rangle^{\tau}+(A^{(3)})^{\tau}\\
&=\sum_{s=1}^{\tau\wedge\cdot}E\left[K_s{{\widetilde{G}_s-G_s}\over{G_{s-1}}}+\widetilde{\cP}^{(3)}_s{{G_s}\over{G_{s-1}}}-\widetilde{\cP}^{(3)}_{s-1}\Big|{\cal{F}}_{s-1}\right].
\end{split}\end{equation*}
This ends the proof of assertion (b), and the proof of the theorem is complete.\end{proof}
\appendix


\end{document}